\documentclass[preprint]{aastex63}
\usepackage{booktabs}
\usepackage{hyperref}
\usepackage{tablefootnote}
\usepackage{verbatim}
\usepackage{todonotes}

\usepackage[utf8]{inputenc}

\begin{document}
\noindent{\sc FERMILAB-PUB-21-469-E}
\vskip 4.7mm

\title{The DES Bright Arcs Survey: Candidate Strongly Lensed Galaxy Systems from the Dark Energy Survey 5,000 Sq. Deg. Footprint}

\correspondingauthor{H.~T.~Diehl} \email{diehl@fnal.gov}

\author[0000-0003-4083-1530]{J.~H.~O'Donnell}
\affiliation{Department of Physics, University of California at Santa Cruz, Santa Cruz, CA 95064, USA}
\author[0000-0002-3908-7313]{R.~D.~Wilkinson}
\affiliation{Department of Physics and Astronomy, Pevensey Building, University of Sussex, Brighton, BN1 9QH, UK}
\author[0000-0002-8357-7467]{H.~T.~Diehl}
\affiliation{Fermi National Accelerator Laboratory, P. O. Box 500, Batavia, IL 60510, USA}
\author[0000-0002-9441-3193]{C.~Aros-Bunster}
\affiliation{Instituto de F\'{i}sica, Pontificia Universidad Cat\'{o}lica de Valpara\'{i}so, Casilla 4059, Valpara\'{i}so, Chile}
\author[0000-0001-8156-0429]{K.~Bechtol}
\affiliation{Physics Department, 2320 Chamberlin Hall, University of Wisconsin-Madison, 1150 University Avenue Madison, WI 53706-1390}
\author[0000-0003-3195-5507]{S.~Birrer}
\affiliation{Kavli Institute for Particle Astrophysics \& Cosmology, P. O. Box 2450, Stanford University, Stanford, CA 94305, USA}
\affiliation{Department of Physics, Stanford University, 382 Via Pueblo Mall, Stanford, CA 94305, USA}
\author[0000-0002-3304-0733]{E.~J.~Buckley-Geer}
\affiliation{Fermi National Accelerator Laboratory, P. O. Box 500, Batavia, IL 60510, USA}
\affiliation{Department of Astronomy and Astrophysics, University of Chicago, Chicago, IL 60637, USA}
\author[0000-0003-3044-5150]{A. Carnero Rosell}
\affiliation{Laborat\'orio Interinstitucional de e-Astronomia - LIneA, Rua Gal. Jos\'e Cristino 77, Rio de Janeiro, RJ - 20921-400, Brazil}
\affiliation{Observat\'orio Nacional, Rua Gal. Jos\'e Cristino 77, Rio de Janeiro, RJ - 20921-400, Brazil}
\author[0000-0002-4802-3194]{M.~Carrasco~Kind}
\affiliation{Department of Astronomy, University of Illinois at Urbana-Champaign, 1002 W. Green Street, Urbana, IL 61801, USA}
\affiliation{Center for Astrophysical Surveys, National Center for Supercomputing Applications, 1205 West Clark St., Urbana, IL 61801, USA}
\author[0000-0002-7731-277X]{L.~N.~da~Costa}
\affiliation{Laborat\'orio Interinstitucional de e-Astronomia - LIneA, Rua Gal. Jos\'e Cristino 77, Rio de Janeiro, RJ - 20921-400, Brazil}
\affiliation{Observat\'orio Nacional, Rua Gal. Jos\'e Cristino 77, Rio de Janeiro, RJ - 20921-400, Brazil}
\author[0000-0001-7282-3864]{S.~J.~Gonzalez Lozano}
\affiliation{Physics Department, 2320 Chamberlin Hall, University of Wisconsin-Madison, 1150 University Avenue Madison, WI 53706-1390}
\author[0000-0002-4588-6517]{R.~A.~Gruendl}
\affiliation{Department of Astronomy, University of Illinois at Urbana-Champaign, 1002 W. Green Street, Urbana, IL 61801, USA}
\affiliation{Center for Astrophysical Surveys, National Center for Supercomputing Applications, 1205 West Clark St., Urbana, IL 61801, USA}
\author[0000-0002-8490-8117]{M. Hilton}
\affiliation{Astrophysics Research Centre, University of KwaZulu-Natal, Westville Campus, Durban 4041, South Africa}
\affiliation{ School of Mathematics, Statistics \& Computer Science, University of KwaZulu-Natal, Westville Campus, Durban 4041, South Africa}
\author[0000-0002-7825-3206]{H. Lin}
\affiliation{Fermi National Accelerator Laboratory, P. O. Box 500, Batavia, IL 60510, USA}
\author[0000-0002-8414-7776]{K.~A.~Lindgren}
\affiliation{Fermi National Accelerator Laboratory, P. O. Box 500, Batavia, IL 60510, USA}
\author{J.~Martin}
\affiliation{Brookhaven National Laboratory, Bldg 510, Upton, NY 11973, USA}
\affiliation{Half Hollow Hills High School East, 50 Vanderbilt Pkwy, Dix Hills, NY 11746, USA}
\author[0000-0001-9186-6042]{A. Pieres}
\affiliation{Laborat\'orio Interinstitucional de e-Astronomia - LIneA, Rua Gal. Jos\'e Cristino 77, Rio de Janeiro, RJ - 20921-400, Brazil}
\affiliation{Observat\'orio Nacional, Rua Gal. Jos\'e Cristino 77, Rio de Janeiro, RJ - 20921-400, Brazil}
\author[0000-0001-9376-3135]{E.~S.~Rykoff}
\affiliation{Kavli Institute for Particle Astrophysics \& Cosmology, P. O. Box 2450, Stanford University, Stanford, CA 94305, USA}
\affiliation{SLAC National Accelerator Laboratory, Menlo Park, CA
94025, USA}
\author[0000-0002-1831-1953]{I.~Sevilla-Noarbe}
\affiliation{Centro de Investigaciones Energ\'eticas, Medioambientales y Tecnol\'ogicas (CIEMAT), Madrid, Spain}
\author[0000-0001-9194-0441]{E.~Sheldon}
\affiliation{Brookhaven National Laboratory, Bldg 510, Upton, NY 11973, USA}
\author[0000-0002-8149-1352]{C.~Sif\'{o}n}
\affiliation{Instituto de F\'{i}sica, Pontificia Universidad Cat\'{o}lica de Valpara\'{i}so, Casilla 4059, Valpara\'{i}so, Chile}
\author[0000-0001-7211-5729]{D.~L.~Tucker}
\affiliation{Fermi National Accelerator Laboratory, P. O. Box 500, Batavia, IL 60510, USA}
\author[0000-0002-9541-2678]{B.~Yanny}
\affiliation{Fermi National Accelerator Laboratory, P. O. Box 500, Batavia, IL 60510, USA}
%
%
\author[0000-0003-1587-3931]{T.~M.~C.~Abbott}
\affiliation{Cerro Tololo Inter-American Observatory, NSF's National Optical-Infrared Astronomy Research Laboratory, Casilla 603, La Serena, Chile}
\author[0000-0001-5679-6747]{M.~Aguena}
\affiliation{Laborat\'orio Interinstitucional de e-Astronomia - LIneA, Rua Gal. Jos\'e Cristino 77, Rio de Janeiro, RJ - 20921-400, Brazil}
\author[0000-0002-7069-7857]{S.~Allam}
\affiliation{Fermi National Accelerator Laboratory, P. O. Box 500, Batavia, IL 60510, USA}
\author{F.~Andrade-Oliveira}
\affiliation{Instituto de F\'{i}sica Te\'orica, Universidade Estadual Paulista, S\~ao Paulo, Brazil}
\affiliation{Laborat\'orio Interinstitucional de e-Astronomia - LIneA, Rua Gal. Jos\'e Cristino 77, Rio de Janeiro, RJ - 20921-400, Brazil}
\author[0000-0002-0609-3987]{J.~Annis}
\affiliation{Fermi National Accelerator Laboratory, P. O. Box 500, Batavia, IL 60510, USA}
\author[0000-0002-3602-3664]{E.~Bertin}
\affiliation{CNRS, UMR 7095, Institut d'Astrophysique de Paris, F-75014, Paris, France}
\affiliation{Sorbonne Universit\'es, UPMC Univ Paris 06, UMR 7095, Institut d'Astrophysique de Paris, F-75014, Paris, France}
\author[0000-0002-8458-5047]{D.~Brooks}
\affiliation{Department of Physics \& Astronomy, University College London, Gower Street, London, WC1E 6BT, UK}
\author[0000-0003-1866-1950]{D.~L.~Burke}
\affiliation{Kavli Institute for Particle Astrophysics \& Cosmology, P. O. Box 2450, Stanford University, Stanford, CA 94305, USA}
\affiliation{SLAC National Accelerator Laboratory, Menlo Park, CA 94025, USA}
\author[0000-0002-3130-0204]{J.~Carretero}
\affiliation{Institut de F\'{\i}sica d'Altes Energies (IFAE), The Barcelona Institute of Science and Technology, Campus UAB, 08193 Bellaterra (Barcelona) Spain}
\author[0000-0001-8158-1449]{M.~Costanzi}
\affiliation{Astronomy Unit, Department of Physics, University of Trieste, via Tiepolo 11, I-34131 Trieste, Italy}
\affiliation{INAF-Osservatorio Astronomico di Trieste, via G. B. Tiepolo 11, I-34143 Trieste, Italy}
\affiliation{Institute for Fundamental Physics of the Universe, Via Beirut 2, 34014 Trieste, Italy}
\author[0000-0001-8318-6813]{J.~De~Vicente}
\affiliation{Centro de Investigaciones Energ\'eticas, Medioambientales y Tecnol\'ogicas (CIEMAT), Madrid, Spain}
\author[0000-0002-0466-3288]{S.~Desai}
\affiliation{Department of Physics, IIT Hyderabad, Kandi, Telangana 502285, India}
\author[0000-0002-8134-9591]{J.~P.~Dietrich}
\affiliation{Faculty of Physics, Ludwig-Maximilians-Universit\"at, Scheinerstr. 1, 81679 Munich, Germany}
\author[0000-0002-1407-4700]{K.~Eckert}
\affiliation{Department of Physics and Astronomy, University of Pennsylvania, Philadelphia, PA 19104, USA}
\author{S.~Everett}
\affiliation{Santa Cruz Institute for Particle Physics, Santa Cruz, CA 95064, USA}
\author[0000-0002-1295-1132]{I.~Ferrero}
\affiliation{Institute of Theoretical Astrophysics, University of Oslo. P.O. Box 1029 Blindern, NO-0315 Oslo, Norway}
\author[0000-0002-2367-5049]{B.~Flaugher}
\affiliation{Fermi National Accelerator Laboratory, P. O. Box 500, Batavia, IL 60510, USA}
\author[0000-0002-1510-5214]{P.~Fosalba}
\affiliation{Institut d'Estudis Espacials de Catalunya (IEEC), 08034 Barcelona, Spain}
\affiliation{Institute of Space Sciences (ICE, CSIC),  Campus UAB, Carrer de Can Magrans, s/n,  08193 Barcelona, Spain}
\author[0000-0003-4079-3263]{J.~Frieman}
\affiliation{Fermi National Accelerator Laboratory, P. O. Box 500, Batavia, IL 60510, USA}
\affiliation{Kavli Institute for Cosmological Physics, University of Chicago, Chicago, IL 60637, USA}
\author[0000-0002-9370-8360]{J.~Garc\'ia-Bellido}
\affiliation{Instituto de Fisica Teorica UAM/CSIC, Universidad Autonoma de Madrid, 28049 Madrid, Spain}
\author[0000-0001-9632-0815]{E.~Gaztanaga}
\affiliation{Institut d'Estudis Espacials de Catalunya (IEEC), 08034 Barcelona, Spain}
\affiliation{Institute of Space Sciences (ICE, CSIC),  Campus UAB, Carrer de Can Magrans, s/n,  08193 Barcelona, Spain}
\author[0000-0001-6942-2736]{D.~W.~Gerdes}
\affiliation{Department of Astronomy, University of Michigan, Ann Arbor, MI 48109, USA}
\affiliation{Department of Physics, University of Michigan, Ann Arbor, MI 48109, USA}
\author[0000-0003-3270-7644]{D.~Gruen}
\affiliation{Faculty of Physics, Ludwig-Maximilians-Universit\"at, Scheinerstr. 1, 81679 Munich, Germany}
\author[0000-0003-3023-8362]{J.~Gschwend}
\affiliation{Laborat\'orio Interinstitucional de e-Astronomia - LIneA, Rua Gal. Jos\'e Cristino 77, Rio de Janeiro, RJ - 20921-400, Brazil}
\affiliation{Observat\'orio Nacional, Rua Gal. Jos\'e Cristino 77, Rio de Janeiro, RJ - 20921-400, Brazil}
\author[0000-0003-2524-5154]{M.~S.~S.~Gill}
\affiliation{Kavli Institute for Particle Astrophysics \& Cosmology, P. O. Box 2450, Stanford University, Stanford, CA 94305, USA}
\author[0000-0003-0825-0517]{G.~Gutierrez}
\affiliation{Fermi National Accelerator Laboratory, P. O. Box 500, Batavia, IL 60510, USA}
\author[0000-0003-2071-9349]{S.~R.~Hinton}
\affiliation{School of Mathematics and Physics, University of Queensland,  Brisbane, QLD 4072, Australia}
\author[0000-0002-9369-4157]{D.~L.~Hollowood}
\affiliation{Santa Cruz Institute for Particle Physics, Santa Cruz, CA 95064, USA}
\author[0000-0002-6550-2023]{K.~Honscheid}
\affiliation{Center for Cosmology and Astro-Particle Physics, The Ohio State University, Columbus, OH 43210, USA}
\affiliation{Department of Physics, The Ohio State University, Columbus, OH 43210, USA}
\author[0000-0001-5160-4486]{D.~J.~James}
\affiliation{Center for Astrophysics $\vert$ Harvard \& Smithsonian, 60 Garden Street, Cambridge, MA 02138, USA}
\author[0000-0001-6089-0365]{T.~Jeltema}
\affiliation{Santa Cruz Institute for Particle Physics, Santa Cruz, CA 95064, USA}
\author[0000-0003-0120-0808]{K.~Kuehn}
\affiliation{Australian Astronomical Optics, Macquarie University, North Ryde, NSW 2113, Australia}
\affiliation{Lowell Observatory, 1400 Mars Hill Rd, Flagstaff, AZ 86001, USA}
\author[0000-0002-1134-9035]{O.~Lahav}
\affiliation{Department of Physics \& Astronomy, University College London, Gower Street, London, WC1E 6BT, UK}
\author[0000-0002-4719-3781]{M.~Lima}
\affiliation{Departamento de F\'isica Matem\'atica, Instituto de F\'isica, Universidade de S\~ao Paulo, CP 66318, S\~ao Paulo, SP, 05314-970, Brazil}
\affiliation{Laborat\'orio Interinstitucional de e-Astronomia - LIneA, Rua Gal. Jos\'e Cristino 77, Rio de Janeiro, RJ - 20921-400, Brazil}
\author[0000-0001-9856-9307]{M.~A.~G.~Maia}
\affiliation{Laborat\'orio Interinstitucional de e-Astronomia - LIneA, Rua Gal. Jos\'e Cristino 77, Rio de Janeiro, RJ - 20921-400, Brazil}
\affiliation{Observat\'orio Nacional, Rua Gal. Jos\'e Cristino 77, Rio de Janeiro, RJ - 20921-400, Brazil}
\author[0000-0003-0710-9474]{J.~L.~Marshall}
\affiliation{George P. and Cynthia Woods Mitchell Institute for Fundamental Physics and Astronomy, and Department of Physics and Astronomy, Texas A\&M University, College Station, TX 77843,  USA}
\author[0000-0002-8873-5065]{P.~Melchior}
\affiliation{Department of Astrophysical Sciences, Princeton University, Peyton Hall, Princeton, NJ 08544, USA}
\author[0000-0002-1372-2534]{F.~Menanteau}
\affiliation{Center for Astrophysical Surveys, National Center for Supercomputing Applications, 1205 West Clark St., Urbana, IL 61801, USA}
\affiliation{Department of Astronomy, University of Illinois at Urbana-Champaign, 1002 W. Green Street, Urbana, IL 61801, USA}
\author[0000-0002-6610-4836]{R.~Miquel}
\affiliation{Instituci\'o Catalana de Recerca i Estudis Avan\c{c}ats, E-08010 Barcelona, Spain}
\affiliation{Institut de F\'{\i}sica d'Altes Energies (IFAE), The Barcelona Institute of Science and Technology, Campus UAB, 08193 Bellaterra (Barcelona) Spain}
\author[0000-0002-7016-5471]{R.~Morgan}
\affiliation{Physics Department, 2320 Chamberlin Hall, University of Wisconsin-Madison, 1150 University Avenue Madison, WI  53706-1390}
\author[0000-0001-6706-8972]{B.~Nord}
\affiliation{Fermi National Accelerator Laboratory, P. O. Box 500, Batavia, IL 60510, USA}
\affiliation{Kavli Institute for Cosmological Physics, University of Chicago, Chicago, IL 60637, USA}
\affiliation{Department of Astronomy and Astrophysics, University of Chicago, Chicago, IL 60637, USA}
\author[0000-0003-2120-1154]{R.~L.~C.~Ogando}
\affiliation{Observat\'orio Nacional, Rua Gal. Jos\'e Cristino 77, Rio de Janeiro, RJ - 20921-400, Brazil}
\author[0000-0003-1339-2683]{F.~Paz-Chinch\'{o}n}
\affiliation{Center for Astrophysical Surveys, National Center for Supercomputing Applications, 1205 West Clark St., Urbana, IL 61801, USA}
\affiliation{Institute of Astronomy, University of Cambridge, Madingley Road, Cambridge CB3 0HA, UK}
\author{M.~E.~S.~Pereira}
\affiliation{Department of Physics, University of Michigan, Ann Arbor, MI 48109, USA}
\affiliation{Hamburger Sternwarte, Universit\"{a}t Hamburg, Gojenbergsweg 112, 21029 Hamburg, Germany}
\author[0000-0002-2598-0514]{A.~A.~Plazas~Malag\'on}
\affiliation{Department of Astrophysical Sciences, Princeton University, Peyton Hall, Princeton, NJ 08544, USA}
\author[0000-0001-6163-1058]{M.~Rodriguez-Monroy}
\affiliation{Centro de Investigaciones Energ\'eticas, Medioambientales y Tecnol\'ogicas (CIEMAT), Madrid, Spain}
\author[0000-0002-9328-879X]{A.~K.~Romer}
\affiliation{Department of Physics and Astronomy, Pevensey Building, University of Sussex, Brighton, BN1 9QH, UK}
\author[0000-0001-5326-3486]{A.~Roodman}
\affiliation{Kavli Institute for Particle Astrophysics \& Cosmology, P. O. Box 2450, Stanford University, Stanford, CA 94305, USA}
\affiliation{SLAC National Accelerator Laboratory, Menlo Park, CA 94025, USA}
\author[0000-0002-9646-8198]{E.~Sanchez}
\affiliation{Centro de Investigaciones Energ\'eticas, Medioambientales y Tecnol\'ogicas (CIEMAT), Madrid, Spain}
\author{V.~Scarpine}
\affiliation{Fermi National Accelerator Laboratory, P. O. Box 500, Batavia, IL 60510, USA}
\author[0000-0001-9504-2059]{M.~Schubnell}
\affiliation{Department of Physics, University of Michigan, Ann Arbor, MI 48109, USA}
\author{S.~Serrano}
\affiliation{Institut d'Estudis Espacials de Catalunya (IEEC), 08034 Barcelona, Spain}
\affiliation{Institute of Space Sciences (ICE, CSIC),  Campus UAB, Carrer de Can Magrans, s/n,  08193 Barcelona, Spain}
\author[0000-0002-3321-1432]{M.~Smith}
\affiliation{School of Physics and Astronomy, University of Southampton,  Southampton, SO17 1BJ, UK}
\author[0000-0002-7047-9358]{E.~Suchyta}
\affiliation{Computer Science and Mathematics Division, Oak Ridge National Laboratory, Oak Ridge, TN 37831}
\author[0000-0002-1488-8552]{M.~E.~C.~Swanson}
\affiliation{Center for Astrophysical Surveys, National Center for Supercomputing Applications, 1205 West Clark St., Urbana, IL 61801, USA}
\author[0000-0003-1704-0781]{G.~Tarle}
\affiliation{Department of Physics, University of Michigan, Ann Arbor, MI 48109, USA}
\author[0000-0002-6325-5671]{D.~Thomas}
\affiliation{Institute of Cosmology and Gravitation, University of Portsmouth, Portsmouth, PO1 3FX, UK}
\author[0000-0001-7836-2261]{C.~To}
\affiliation{Department of Physics, Stanford University, 382 Via Pueblo Mall, Stanford, CA 94305, USA}
\affiliation{Kavli Institute for Particle Astrophysics \& Cosmology, P. O. Box 2450, Stanford University, Stanford, CA 94305, USA}
\affiliation{SLAC National Accelerator Laboratory, Menlo Park, CA 94025, USA}
\author{T.~N.~Varga}
\affiliation{Max Planck Institute for Extraterrestrial Physics, Giessenbachstrasse, 85748 Garching, Germany}
\affiliation{Universit\"ats-Sternwarte, Fakult\"at f\"ur Physik, Ludwig-Maximilians Universit\"at M\"unchen, Scheinerstr. 1, 81679 M\"unchen, Germany}
%
%
\collaboration{1000}{(DES Collaboration)}


\begin{abstract}
We report the combined results of eight searches for strong gravitational lens systems in the full 5,000 sq. deg. of Dark Energy Survey (DES) observations.
The observations accumulated by the end of the third observing season fully covered the DES  footprint in 5 filters (grizY), with an $i-$band limiting magnitude  $(at 10\sigma)$ of 23.44. 
In four searches, a list of potential candidates was identified using a color and magnitude selection from the object catalogs created from the first three observing seasons. Three other searches were conducted at the locations of previously identified galaxy clusters. Cutout images of potential candidates were then visually scanned using an object viewer. An additional set of candidates came from a data-quality check of a subset of the color-coadd ``tiles" created from the full DES six-season data set.  A short list of the most promising strong lens candidates was then numerically ranked according to whether or not we judged them to be bona fide strong gravitational lens systems. These searches discovered a diverse set of 247 strong lens candidate systems, of which 81 are identified for the first time. We provide the coordinates, magnitudes, and photometric properties of the lens and source objects, and an estimate of the Einstein radius for 81 new systems and 166 previously reported. This catalog will be of use for selecting interesting systems for detailed follow-up, studies of galaxy cluster and group mass profiles, as well as a training/validation set for automated strong lens searches.
\end{abstract}


\section{Introduction}

Gravitational lensing occurs when the gravitational field of a massive foreground object, the ``lens'', deflects the light of a more distant object, the  ``source".  If the lens is massive enough, and the source, lens, and observer sufficiently collinear and distant from one another, the observed image of the source object can be distorted into rings, arcs, or multiple images. This effect is known as strong gravitational lensing (SL), and has become a powerful and popular tool for studying a wide variety of extragalactic physics and cosmology.

Though strong gravitational lensing of very distant objects by foreground galaxies was predicted~\citep{zwicky} four decades prior, the first discovery of a strongly lensed object was that of QSO 0957+561~\citep{qso0957}, a double image of a single quasar at redshift $z = 1.4$. Not long after, it was realized that the long, narrow, curved features bent around the centers of galaxy clusters were actually SL systems as well~\citep{soucail1987, lynds1989,  grossman1989}.  In the subsequent 40-year interval, the count of candidate strongly lensed galaxies has expanded to more than 2000.  At the same time, so has their scientific interest.

Strong gravitational lensing provides new opportunities to study astrophysics and the physics of dark matter. Because the surface brightness of a source galaxy is unchanged by lensing, magnification of the source provides amplification of the image flux.  Colloquially named ``Einstein's Telescopes"~\citep{gates2009}, strong-lens systems enable studies of details of distant galaxies that would otherwise be unresolved or too faint. For example, faint lensed sources with relatively large redshifts enable studies of the initial mass-to-light ratio, star formation and metallicity in young galaxies~\citep[e.g.][]{bayliss2014, leier2016}. Studies of the lens systems, whether galaxies, galaxy groups, or galaxy clusters,  provide information on their total mass and mass profiles, including baryon and dark matter~\citep[e.g.][]{koopmans2009, wiesner,  treuellis2014, newman2015}, and explore the connection between these characteristics and the occurrence of strong lensing~\citep{robertson2020, fox2021, sonnenfeld2021}. Strong lens systems are now being used to study galaxy and cluster substructures and to, in turn, constrain the interaction physics of dark matter~\citep{meneghetti2020, gilman2021}.

Strong gravitational lensing enables studies of cosmology. Studies of characteristics of ensembles of strong lensing systems can be used to constrain $\Omega_M$ in wCDM and $\Lambda$CDM~\citep{leaf2018}, albeit loosely, with samples of O(100) confirmed lensing systems.  Studies of gravitationally lensed time-varying systems~\citep{refsdal, bland, birrer, oguri2019}, such as lensed quasars~\citep{pls1, suyu} or supernovae~\citep{grillo}, are used to constrain the expansion history between source, lens, and observer.\footnote{Note that our SL searches could identify lensed quasars, but it is unlikely that we would have identified short duration transients, such as SN, in our wide-field images.} 
Individual measurement of strongly-lens quad quasars systems provide~\citep{h0licowV, shajib2020} constraints on $H_0$ with about $4\%$ precision, and combinations~\citep{h0licow1, h0licow13} are now in mild tension, or not~\citep{Birrer2020}, with early Universe probes, depending~\citep{BT2021} on assumptions about the lens' radial mass profiles. Lens systems with multiple sources at differing redshifts can provide~\citep{linkpierce,gavazzi08,collett2012} complementary information about the expansion history, independent of the Hubble constant. Constraints on $\Omega_M$ and the dark energy equation of state, $w$, are at about the $20\%$ level from individual lensing systems~\citep{jullo10, collett2014}. 
{{In these systems it can be critical to accurately model massive lens substructures to obtain unbiased cosmological constraints~\citep{daloisio2011}.}}



The count of candidate SL systems has increased rapidly with the advent of deep wide-field surveys with dedicated SL searches. In the CFHTLS Strong Lensing Legacy Survey~\citep{Cabanac.Alard.ea2007}, 54 systems with promising lenses were identified by the {\sc ARCFINDER} algorithm~\citep{more2012} in 150 sq. deg., 49 confirmed strong lens systems identified with {\sc RingFinder}~\citep{gavazzi2014}, and 29 promising and 59 total~\citep{sw2-2016} from a crowdsourcing~\citep{sw1} effort. CFHTLS imaging yielded 16 more probable to definite SL systems using neural network-based  searches~\citep{Jacobs.Glazebrook.ea2017}.  The Survey of Graviationally-lensed Objects in HSC Imaging (SuGoHI, \cite{Sonnenfeld.Chan.ea2018}) used three different methods, including {\sc YattaLens}, to search the Hyper Suprime-Cam Subaru Strategic Program (HSC SSP) images. The program yielded 333 galaxy-galaxy SL candidates from an area of 442 deg$^2$.   The Kilo Degree Survey images were searched using machine-learning (ML) classifiers that discovered 227 high grade SL candidates, plus an additional 200 candidates discovered serendipitously~\citep{Petrillo.Tortora.ea2017, Petrillo.Tortora.ea2019, Li.Napolitano.ea2020}.
\cite{Canameras.Schuldt.ea2020} discovered 330 new SL systems in the 30,000 sq. deg. Pan-Starrs $3\pi$ Survey~\citep{chambers2016}.
The 1.64 deg$^2$ Hubble Space Telescope (HST) {\sc COSMOS} survey field yielded 67 galaxy-galaxy lens candidates~\citep{faure2008}.  HST was also used to confirm a total of 110 SL systems
identified~\citep{Bolton.Burles.ea2008, Shu.Brownstein.ea2017} in the Sloan Digital Sky Survey (SDSS) spectroscopic data. SDSS also yielded 19 confirmed systems to the Sloan Bright Arcs Survey~\citep{sbas-8oclock,sbas-kubo1,clone,sbas-diehl,sbas-kubo2}, and more than 30 confirmed and 50 additional candidate lenses to the CASSOWARY survey~\citep{belloch2009,pettini2010,Stark2013}. Searches of galaxy clusters identified in SDSS yielded 16 strong lens systems with $> 10$\arcsec \ radius and 21 additional SL candidates~\citep{hennawi2008}, 68 giant arcs in \cite{wen2011}, ten SL systems in \cite{furlanetto2013}, and 37 more in \cite{sharon2020}. More recently, a search of galaxy clusters in HSC imaging \citep{Jaelani2020} revealed more than 600 candidate strong lens systems, of which 8 are confirmed spectroscopically. The eBOSS spectroscopic data produced~\citep{Talbot.Brownstein.ea2021} 838 likely galaxy-galaxy SL systems with Einstein radius $\lesssim 1\arcsec$. The DESI Legacy Imaging Surveys team used 14,000 sq. degs. of images from several instruments, including the Dark Energy Camera~\citep{decampaper}, 
in combination with a neural network {{trained on real SL systems}, to identify 1,545 candidate systems~\citep{Huang.Storfer.ea2020, Huang.Storfer.ea2021}.}

Galaxy clusters, among the most massive objects in the universe, are ideal candidates for strong lenses, and optical observations of galaxy cluster catalogs have yielded many SL systems. Galaxy cluster catalogs selected via optical observations, the Sunyaev-Zel'dovich effect~\citep{SZE}, or as extended high-flux x-ray sources have been used. Follow-up of South Pole Telescope (SPT)~\citep{reichardt2013} detections yielded 34 new and 10 previously
known SL systems in \cite{SPTBleem}, and 44 new SL systems in \cite{Bleem2020}. Follow-up of Atacama Cosmology Telescope galaxy cluster identifications~\citep{hilton2021} yielded 67 new candidates including many in the DES  footprint that are characterized here for the first time.


The DES data have proven to be a productive source of SL candidate systems.  Searches for lensed quasars have yielded four spectroscopically confirmed systems~\citep{agnello2015,ostrovsky17,lin17}, two new quad SL systems and 96 more quasar lens candidates ripe for spectroscopic follow-up~\citep{agnello2019}. Spectroscopic observations by the STRIDES collaboration confirmed 10 new lensed quasars and 10 quasar pairs~\citep{Lemon2020}.
Visual scanning of $~400,000$  color cutout images at coordinates selected, using techniques similar to those of this paper, from the $\sim$250 sq. deg. Science Verification (SV) and $\sim$1,800 sq. deg. Y1 catalogs yielded 348 new SL candidates in \cite{diehlsvy1}, as well as one serendipitous SL discovery~\citep{bettinelli2016}.   The convolutional neural network-based search technique described in \cite{Jacobs.Glazebrook.ea2017} contributed~\citep{Jacobs.Collett.ea2019_alt, Jacobs.Collett.ea2019a} 485 ranked candidates in the full $\sim$5,000 sq. deg. DES footprint, principally of the galaxy-galaxy lens configuration. The latter paper also included 26 additional SL candidates identified by visual scanning of $\sim$54,000 potential lensing targets identified using procedures similar to those of \cite{diehlsvy1}. Spectroscopic confirmation of 17 DES systems is described in \cite{Nord.Buckley-Geer.ea2016, lin17, collett2017, nord2020}.

The recent competitive success of automated and machine learning methods for identifying strong lens candidates is evident from a simple comparison of the numbers of candidates in these references.
Neural network and ML algorithm searches for SL systems have advanced because of increased attention to the technique's power and capabilities. Advances in computer vision have given rise to automated searches for strong lens candidates, and for classification, deblending, and modelling of strong lens candidates
~\citep{alard06, Arcfinder, kubomethod, furlanetto2013b, rjoseph, PF2016, PICS, xu2016,  Bom2017,  lanusse2017, SLchallenge, plazas2020}.
 Nonetheless, these ML searches continue to benefit from visual scanning and ranking of candidate systems. Furthermore, continued efforts to attain large samples of real lenses with a variety of morphologies is important for training and validation of new automated lens-finding algorithms, particularly for identification of group and cluster-scale SL systems, where ML has been less successful.




In this paper we report 247 strong gravitational lens candidates from searches of the full 5,000 sq. deg. DES footprint, including 81 previously unreported candidates. These lenses constitute a diverse set of  primarily galaxy group- and cluster-scale systems,  with both red and blue sources.  We provide details of the locations (RA \& DEC), and g,r,i,z,Y band magnitudes of source and lens objects, as well as Einstein radii and a rank, quantifying the confidence level of the selection, for each SL candidate system.
This paper is organized as follows.  In $\S 2$ we describe the Dark Energy Survey observations and catalogs.   In $\S 3$ we describe our strong gravitational lens search procedures.   In $\S 4$ we describe the results from the searches and provide the properties of the candidate lens systems. We highlight some of the systems that have notable properties.  Finally, in $\S 5$ we summarize and discuss our results.

\color{black}

\section{Dark Energy Survey Imaging Data}
\label{sec:des-data}
The Dark Energy Survey has completed a six season program that imaged 5,100 sq. deg. of the southern galactic cap using the Dark Energy Camera (DECam)~\citep{decampaper}, which is operated on the 4m Victor M. Blanco Telescope at Cerro Tololo Interamerican Observatory (CTIO) near La Serena, Chile.
Following a ``Science Verification" period~\citep{lahav}, DES data collection typically occurred from mid-to-late August to mid-February, starting on August 31, 2013 and ending on January 9, 2019.  Over the six year survey, we accumulated ten high-quality, 90-second duration observations of the full survey field in each of the four filters $g$, $r$, $i$, and $z$-bands, and six to ten high-quality observations of 45 or 90 seconds length for the $Y$-band (totalling 450 seconds).   In addition to the wide-field survey, DES performed a time-domain supernova survey during the same time period, visiting ten fields  in the $g$, $r$, $i$, and $z$-band filters with an approximately weekly cadence and at much greater depth~\citep{kessler} than the wide-field survey. Details of the observation strategy, survey operations, and the survey progress for each of the seasons are available~\citep{neilsen2019,opsy1y3,opsy4y5,opsy6}.  Most of the searches described in this paper used as a starting point the object catalogs and images from the first three seasons of observations.  Much of the information that we provide about the candidate strong lens systems comes from the full six-season data set. When we refer to ``Y1"~\citep{drlicaY1}, ``Y3"~\citep{Y3DR1}, and ``Y6"~\citep{Y6DR2} data sets, we mean to include all the preceding seasons of observations. For instance, Y3 refers to the data from the first three seasons.

The observations were processed by the Dark Energy Survey Data Management (DESDM) system~\citep{mohr,pipeline} in three pipelined stages: single epoch ``detrending", photometric calibration, and coaddition.  The detrending operation removes the instrumental signature, subtracts the sky background, and removes artifacts such as cosmic rays from the individual exposures. The resulting images are in FITS-formatted files with an inverse-variance weight
(WGT), and a mask of bad pixels (MSK). Single epoch catalogs were produced using \textsc{PSFex}~\citep{psfex} and \textsc{SExtractor}~\citep{sextractor}.
Astrometric calibration is performed by matching bright stars on each exposure to reference stellar catalogs (2MASS~\citep{2MASS} for Y3, GAIA DR2~\citep{GAIADR2} for Y6) using \textsc{scamp}~\citep{scamp}. Magitude zeropoints for each CCD on each exposure are determined using the ``Forward Global Calibration Module"~\citep{fgcm}.  Finally, exposures in each filter were coadded using \textsc{SWarp}~\citep{swarp} in 10,000 by 10,000 pixel ``tiles" 0.72 degrees on a side.  \textsc{SExtractor} was then run on a weighted combination of coadded $r+i+z$ ``detection tiles" to form catalogs of objects, and then rerun on the individual filter bands at the location of the detected objects.

The separation or ``deblending" of closely positioned (or even overlapping) objects is a challenge where the goal is to balance completeness against the spurious separation of features within a single galaxy. The deblending was performed using the detection images. 
For Y6, the SourceExtractor source detection threshold was lowered from $10\sigma$ (in Y3) to $5\sigma$ to detect fainter sources; this was accompanied by a reoptimization of the deblending parameters. As a result, objects are sometimes detected and/or deblended better in one data release than they are in the other.  The absolute photometric calibration for both Y3 and Y6 is tied to the spectrophotometric Hubble CALSPEC standard star C26202~\citep{Bohlin}. Other differences between the Y3 and Y6 processing are described in~\citealp{Y6DR2}.  All of the previous seasons' observations were reprocessed to produce the new data sets. Each time, object catalogs containing the list of objects, their shapes, and their astrometric and photometric properties were calculated for each coadd tile.   Unless noted otherwise, the \textsc{SExtractor} MAG\_AUTO magnitudes are the primary measures of coadd flux used in further analysis. 

Photometric redshifts (photo-z's) are derived by the DNF algorithm \cite{DNF16}. The lens photo-z’s are reasonably well-estimated, given that our lens samples consist predominantly of red galaxies, which have strong 4000 Å break features that yield better photo-z measurements. However, we caution that our sources, which are typically fainter blue objects, will have photo-z’s that are subject to larger uncertainties and systematic errors.

\begin{figure}[ht]
\begin{center}
\includegraphics[scale=0.5]{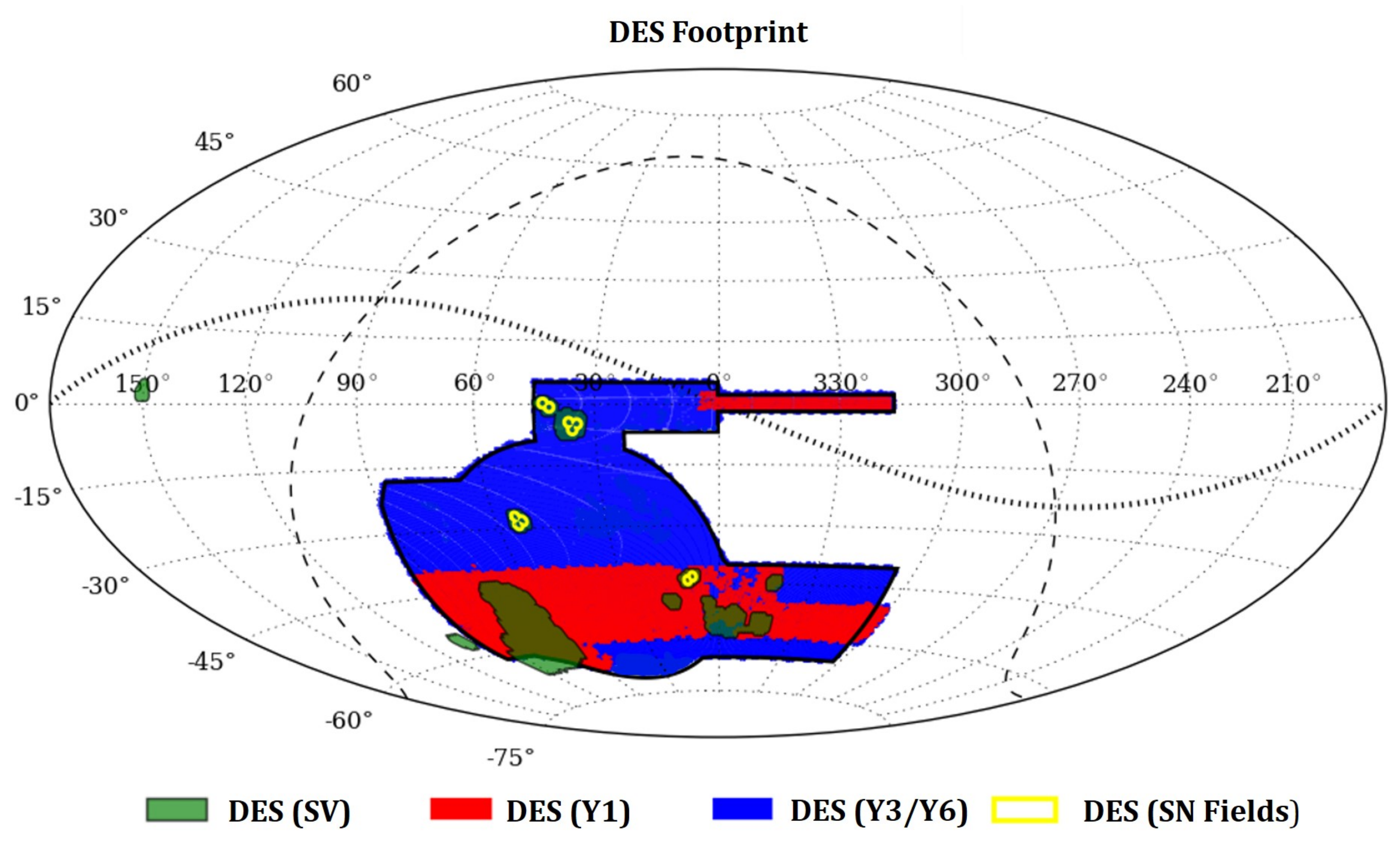}
\caption{
The footprint of the Dark Energy Survey. This result is based on searches of the DES Y3/Y6 footprint, outlined in black. Note that the DES Y3 footprint included additional observations of the Y1 and SV fields, and the the Y6 footprint included additional observations in all of those.}
\label{fig:footprint}
\end{center}
\end{figure}

The DES Y1 and Y3/Y6 observing footprints are described in~\cite{lahav} and shown in Figure~\ref{fig:footprint}. Table~\ref{tab:y1y3y6} characterizes these data sets for comparative purposes. Additional information about the ``Science Verification" (SVA1) data can be found in \cite{diehlsvy1}.

\begin{table}[ht]
\begin{center}
\caption{ \label{tab:y1y3y6} Comparison between Y1, Y3, and Y6 catalogs. Magnitude limits are given for detections with S$/$N $\ge 10$. }
\begin{tabular}{|c|c|c|c|} \hline \hline
 Parameter      &  Y1        & Y3        &   Y6     \\ \hline \hline
Sky Coverage, grizY intersection (deg$^2$)
                & 1793       & 5186      & 4913     \\ \hline
Objects (Galaxies/Stars)
                & 137M Total & 310M/80M  & 543M/145M \\ \hline
i-band Uniformity (typical \# of exp.)
                & 3 to 4     & 4 to 6    & 8 to 10   \\ \hline
i-band Single-epoch Median PSF FWHM (\arcsec)
                &  0.97      & 0.88      & 0.88     \\ \hline
 Coadd Median Astrometric Relative Precision (ang. dist., mas)
                &  25        & 30        &   27     \\ \hline
 i-band Coadd Mag. Lim. (MAG\_AUTO)
                & 22.5       & 22.5      & 23.1     \\ \hline
 i-band Coadd Mag. Lim. ({\sc MAG\_APER\_4}, 1.95{\arcsec} diam.)
                & 22.9       & 23.4      & 23.8     \\ \hline \hline
\end{tabular}
\end{center}
\end{table}

\section{Gravitational Lens Candidate Search Procedures}
\begin{figure}
\begin{center}
\includegraphics[width=1.0\linewidth]{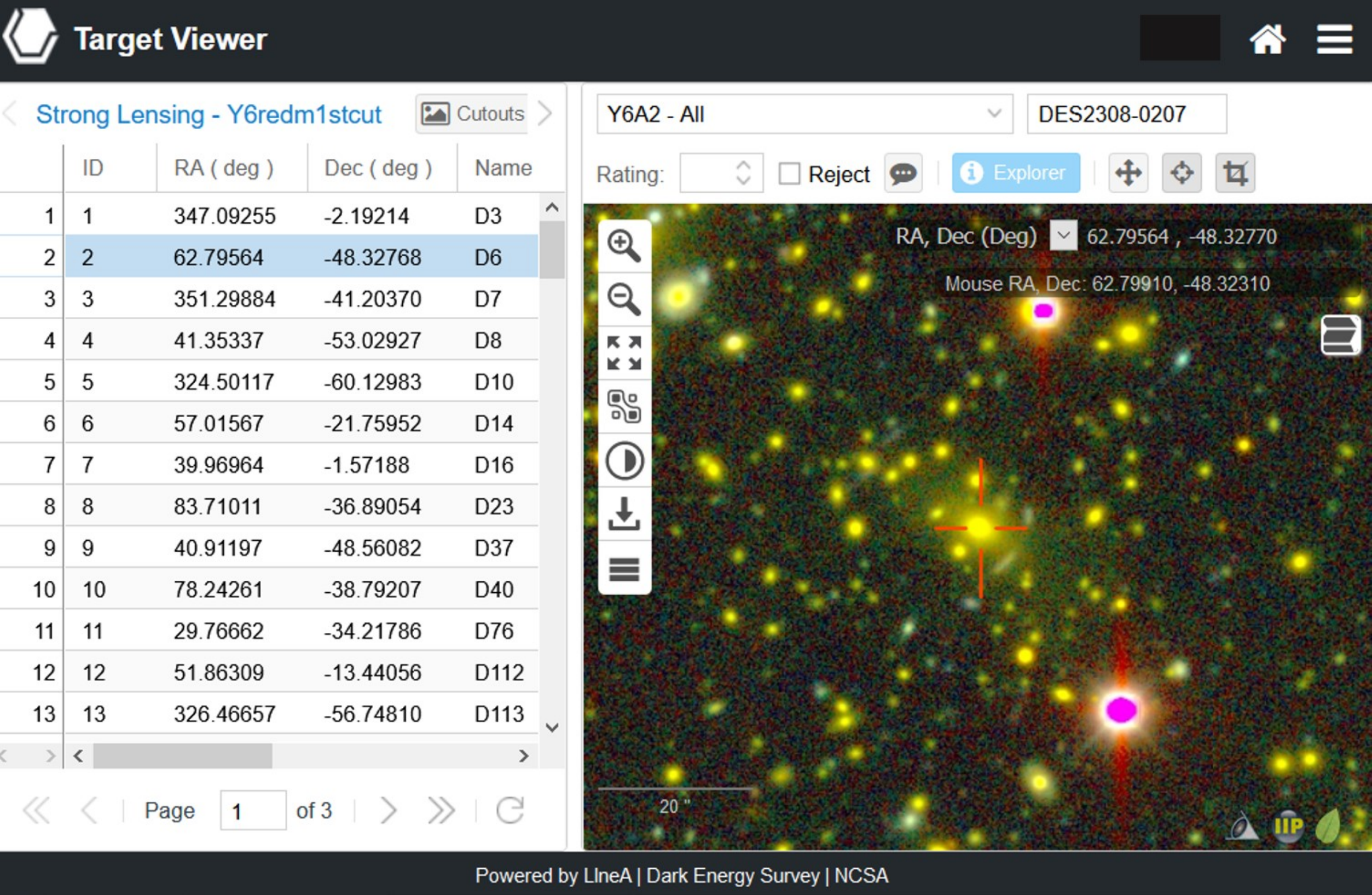}
\caption{An example of an SL candidate shown in the NCSA Target Viewer. The left hand panel is a list of cutouts and characteristics, such as the RA and DEC. On the right side of the viewer is the optical image as well as some of the controls for manipulating the viewer. One cutout is displayed at a time, with an adjustable size of about 2\arcmin\, on a side. In addition, the Target Viewer can access DES object catalogs and encircle the identified objects (not shown). Each circle can display useful information about the object, including the OBJECTID, which provides reference to the object in the respective data release.}
\label{fig:TargetViewer}
\end{center}
\end{figure}

We completed eight different searches, described below, to identify candidate strong lens systems using the Y3 and Y6 data. One of the searches was a by-product of the Y6 data quality effort. The seven other searches involved identifying and visually scanning a list of potential SL systems.  While the origin of the lists was different for each search (those details are described below), the rest of the procedure is in common.  The lists were loaded into the ``DES Science Portal"~\citep{Y6DR2}, which includes ``Target Viewer", a tool for visualizing the DES fields that can also provide catalog information about the objects. The Target Viewer produced 5-color ($g$, $r$, $i$, $z$, and Y-band) cutout images, roughly 2\arcmin \ across,  with one system centered on each page.  Cutouts were scanned by either scientists with experience identifying strong lens systems or students trained to do so.  Each page required only a few seconds to scan.   Figure~\ref{fig:TargetViewer} shows a sample page as seen on the Target Viewer. Potential SL candidates were identified by the occurrence of an apparent arc, or a pattern of arc-like knots or objects suggestive of an instance of strong lensing. It was not required that the potential sources or lenses that we identified were part of the selection that caused the cutout to be made in the first place.  Interesting candidates were flagged for further evaluation. Some bright or particularly interesting candidates were immediately designated for further study.

Each search discovered new SL candidate systems. Some new systems were found by more than one of our searches, as well as some that had previously been discovered. After each search was completed,  a short list of systems identified as candidates was compiled and all re-ranked by a team of five scientists. Each person assigned a score of 0 to 2 to each system - 0 points if the system was thought to not be a SL candidate, 1 point if it might be, and 2 points if the system was expected to be an instance of strong lensing. Scores from all judges were summed to produce a ``rank" between 0 and 10. Systems with a rank of at least 3 were taken as the final list for this paper.  The candidate rankings of 3 to 10 span the range from ``possible'' to ``probable'' to ``definite'' strong  lens systems with rankings  consistent with those used in the Master Lens Database~\citep{MLD} and other graded samples of strong lens candidates. 


\color{black}

\subsection{Blue Near Anything ``Knot" Searches}
We searched the DES Y3 catalogs for SL candidates using a ``Blue Near Anything" (BNA) algorithm, originally motivated by~\cite{kubik} and used extensively by~\cite{diehlsvy1}. This algorithm is aimed at identifying star-forming Lyman break galaxies and Lyman-$\alpha$ emitting galaxies strongly lensed by massive luminous red galaxies (LRGS). There were two such searches.

\begin{figure}
\begin{center}
\includegraphics[width=0.5\linewidth]{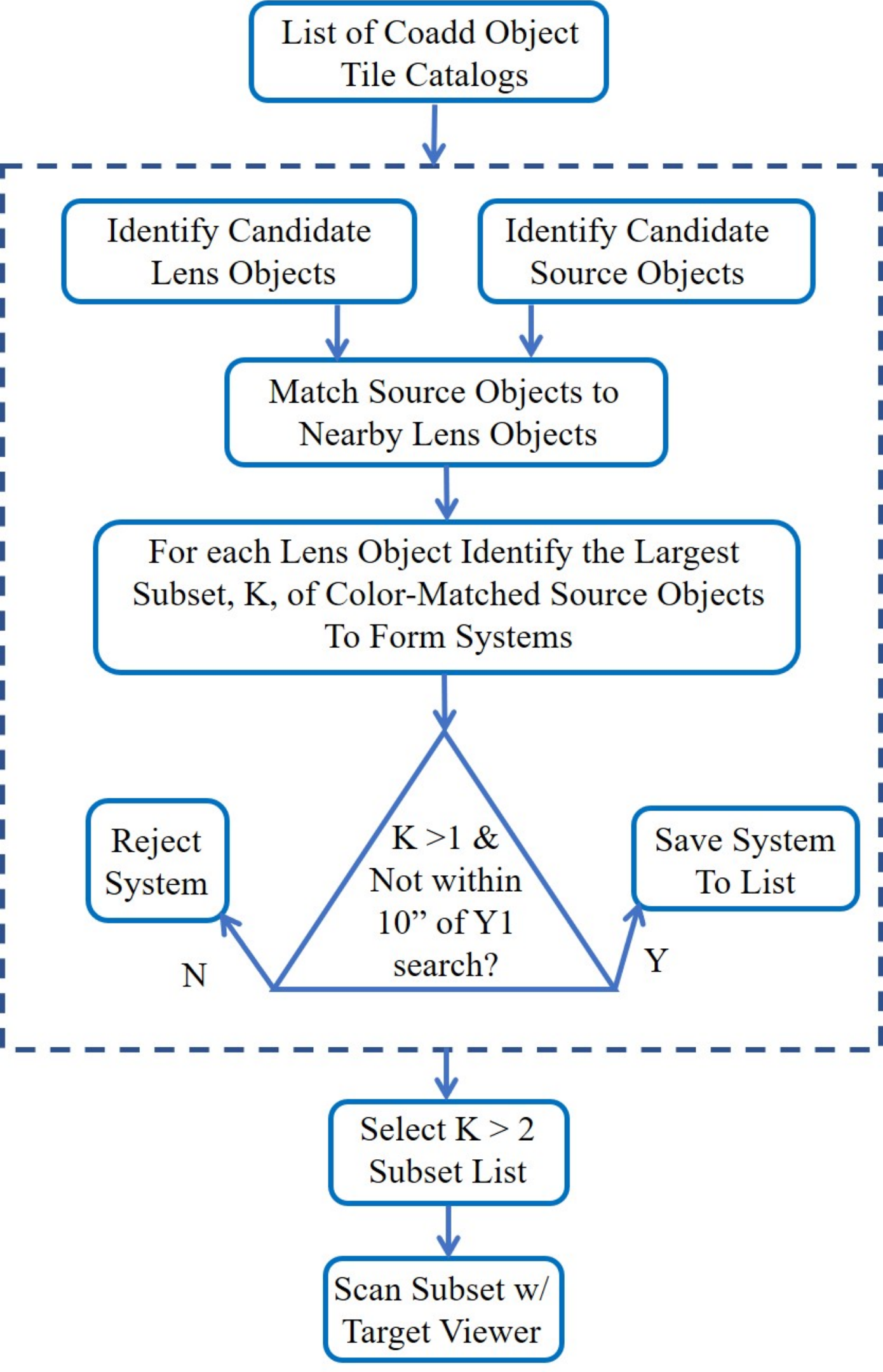}
\caption{Typical flowchart for the ``Blue Near Anything" and ``Red Near Anything" search algorithms.  The box with the dotted outline is a single computer program. At the last step within the program, `Y' and `N' stand for `Yes' and `No'. The ``K $>$ 2" subset selection was done on 3 of 4 of these searches.}
\label{fig:FlowChart}
\end{center}
\end{figure}

The first was performed on the single coadd tile object catalogs  and is illustrated in Figure~\ref{fig:FlowChart}. First, a list of candidate lens galaxies is created.  The criteria for a galaxy to be in the list of possible lenses are that at least one of the $r$-band, $i$-band, or $z$-band magnitudes is less than 21.5.  Selection criteria on \textsc{Sextractor} outputs removed galaxies that were faint, objects that were not well deblended, objects that are likely to be stars, and artifacts left over from objects with saturated pixels.
{{In each \textonehalf \ sq. deg. tile, this search typically identified 5,000 to 7,000 candidate lens galaxies.}}
Next, we formed a list of source candidates.
The criteria for an object to be in the list of possible sources are that at least one of the magnitudes for the $g$-band, $r$-band, or $i$-band must also be less than 21.5 and that the object was not poorly deblended or contained saturated pixels. We did not apply star-galaxy separation in order to preserve strongly lensed quasars which appear star-like~\citep{slreed2015}. Blue-colored source candidates were selected by requiring that $g-r <1.0$ and  $r-i <1.0$. There were typically 2,000 to 3,000 candidate source objects per tile. Next, for each object in the lens list we identified the objects in the source list that were within $8\arcsec$ of the lens. Then, we identified the largest set of those sources, associated with a given lens candidate object, that each had a similar color, where the ``similar" requirement was that  $|\Delta(g-r)|$ and $|\Delta(r-i)|$ both be less than 0.25 magnitudes. 
This algorithm predominantly finds  blue-colored source galaxies lensed by red galaxies or galaxy clusters, and we refer to it as the BNA~2+ selection.  After we identified the list of candidates on each tile, we removed those candidates that were within 10\arcsec \ of a position that we had previously scanned in the Y1 BNA search~\citep{diehlsvy1}.     There were 247,076 systems that had 2 or more matched source objects in the 10345 tile catalogs. We performed a visual scan of the 30,319 systems that had 3 or more matched source objects.  This yielded a short list of 95 SL candidates.

The second BNA search was similar, but it was aimed at identifying those bright SL candidates that we might have missed by scanning only those systems with three or more blue knots. For this ``Bright Blue Near Anything" (BBNA) selection, we raised the lens galaxy magnitude threshold to 20.5 and the source galaxy threshold to 20.0 and reran the algorithm. 
We performed a visual scan of the 18,985 resulting systems with two or more matched source objects. This yielded a short list of 8 additional BBNA SL candidates, all of which received a rank of 3 or more.

\subsection{Red Near Anything 3+ ``Knot" Search} 
The motivation for this search, refereed to as ``Red Near Anything" (RNA), was to discover SL systems with red-colored sources.  It was similar to the RNA search used extensively for Y1 in~\cite{diehlsvy1}, and similar to the BNA search described above except as noted. The lens candidate selection criteria required that any of $r,i,z < 22.0$.   In both iterations, there were two selection criteria for the source lists. The first was that any of $r,i,z < 22.0$, that $g > 23$, and that $g-r>0$ and $r-i>0$. The second was that any of $r,i,z < 22.0$, that $g > 23$ and $r > 23$, and that $r-i>0$ and $i-z>0$. Again, a list of source candidates was matched against a list of lens candidates with an 8\arcsec\ maximum radius. Next, for each system we found the largest set of matching sources for which $|\Delta (g-r)|<0.25$ and $|\Delta (r-i)|<0.25$. After eliminating candidates within 10\arcsec of a location scanned in~\cite{diehlsvy1}, there were 439,077 candidates. We visually scanned 30,994 targets with 3 or more source ``knots", short-listing 110 for further study.

We carried out the second iteration of this campaign on the Y3 data after we realized that the color-matching selection criterion $|\Delta (g-r)|<0.25$ would eliminate source objects that were $g$ and $r$-band dropouts. We reran the algorithm, this time requiring that $|\Delta (r-i)|<0.25$ and $|\Delta (i-z)|<0.25$. A total of 501,184 candidate systems remained after removing locations we had checked in~\cite{diehlsvy1}. There were 20,203 targets with 3 or more source knots. We visually scanned 5,100 of them, finding four systems for the short list. Due to its low productivity, this search (RNA2) was abandoned before we scanned all the targets. All four of these had a grade of 3 or more.

\subsection{An Undirected Search of DES Y6 Tiles}
\begin{figure}
\begin{center}
\includegraphics[width=0.8\linewidth]{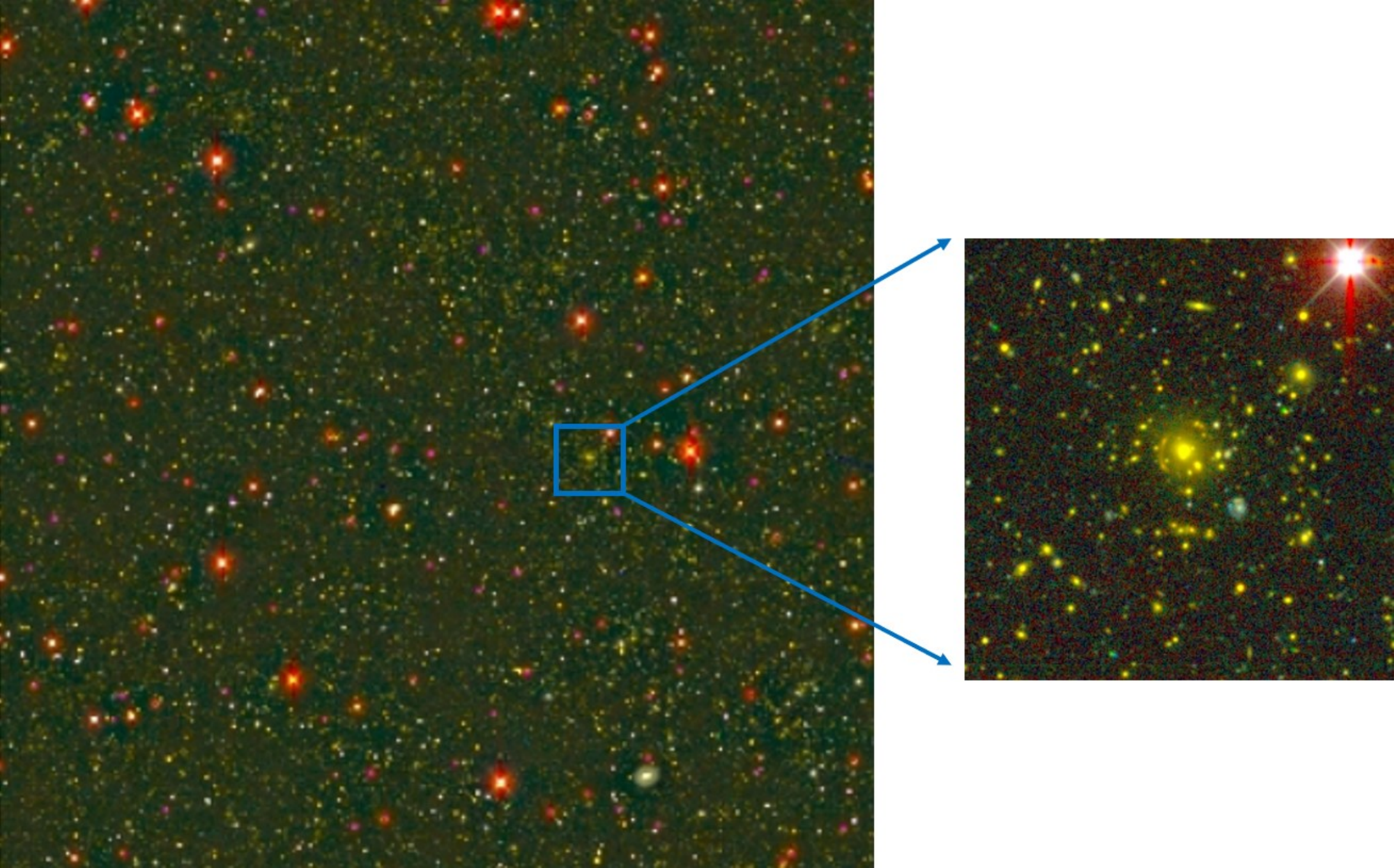}
\caption{In the visual data quality test, we first displayed the full tile of DES data, 0.72 deg on each side (left). Then we zoomed-in a factor of 50 to 100 (in area) at roughly a dozen places on each tile, looking for problems noticeable in a cutout roughly 5\arcmin \ on a side. If we noticed a potential SL candidate we recorded its position. The right-side is about 1/4 of our typical zoomed-in area, but at the scale that we viewed it.
}
\label{fig:TileScan}
\end{center}
\end{figure}

As part of the Y6A1 DES data release, we performed a data quality check comprising a visual scan of a subset of the 10,167 1/2-sq. deg. coadd tiles. The scanning took place in two stages. Initially 2,377 tiles were scanned in RA-order starting with RA's close to 0 hrs. Later, we repeated the process on the 1,708 tiles that had not had any targets in the BNA or RNA SL searches. We were looking for obvious quality defects such as the inclusion of problematic single exposures into the coadds. We identified coadds that included exposures that had been taken during an earthquake, some that had a trail from the reflections off the International Space Station, and others that had a subtle astrometric failures that resulted in a ``doubling-up" of the objects in one filter band. All of these bad exposures were removed from the data set, and coadds were remade without them. This visual scan procedure was as follows. First, we displayed the full coadd tile at a scale that permitted us to notice macroscopic problems,  but not with a resolution to notice SL systems. Then we ``zoomed-in" at a dozen or more different places on the coadd tile, looking for problems that could only be seen with finer resolution. See Figure~\ref{fig:TileScan}. In this process we viewed only a fraction of the area of the tile, perhaps one-tenth, with sufficient resolution to spot a candidate SL system. We recorded 44 candidates for the short list. Seventeen of these were graded as rank 3 or more.

\subsection{Scans of Galaxy Cluster Catalogs}
\label{sec3.4}
\begin{figure}
\begin{center}
\includegraphics[scale=0.8]{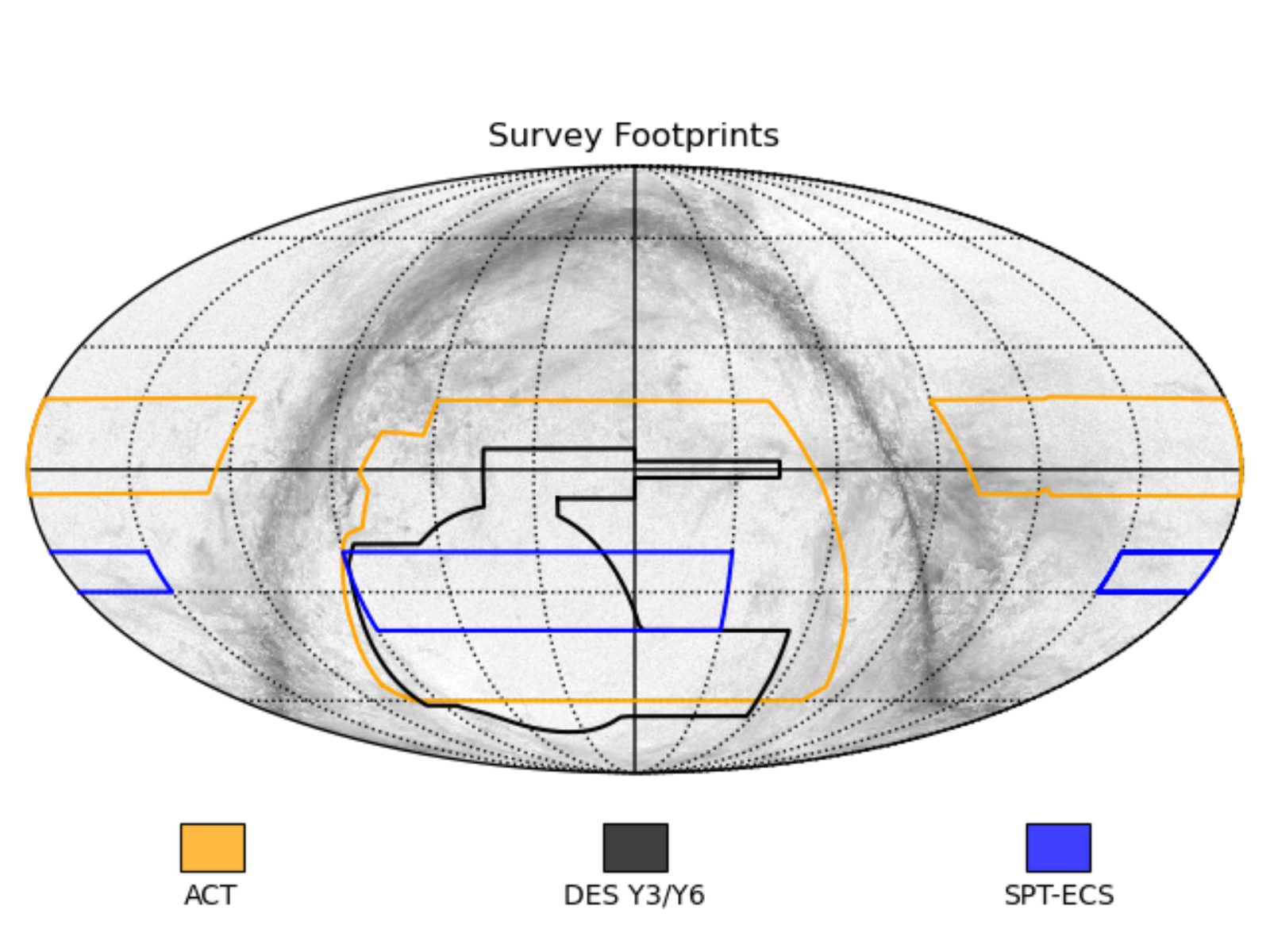}
\caption{
The footprints of the Dark Energy Survey and other cluster catalogs referenced in this work. The DES redMaPPer catalogs span the DES footprint, outlined in black in the center. Searches were also conducted based on galaxy cluster catalogs from the SPTPol-ECS and ACT, whose footprints are shown in blue and orange, respectively. Background dust map obtained from Planck Data Release 2. The overlap between the DES and ACT cluster footprints is approximately 4770 sq. degs., and the overlap between the SPT and DES footprints is approximately 1510 sq. degs. The intersection of all three footprints is approximately 1470 sq. degs.
}
\label{fig:clustersfootprint}
\end{center}
\end{figure}

Galaxy clusters are among the most massive structures in the Universe, and form high-magnification regions where SL systems are likely to appear. We searched for SL systems in three catalogs of galaxy clusters: the optically-selected DES Y3 redMaPPer  catalog, and two cluster catalogs selected via the Sunyaev-Zel'dovich effect \citep{SZE}: the Atacama Cosmology Telescope (ACT) cluster catalog~\citep{hilton2021}, and the SPTPol-Extended Cluster Survey catalog~\citep{Bleem2020}. The overlap of all cluster catalog footprints is shown in Fig~\ref{fig:clustersfootprint}.\footnote{The dust map was obtained from Planck at http://irsa.ipac.caltech.edu/data/Planck/release$\_$2/all-sky-maps/maps/component-maps/foregrounds/COM$\_$CompMap$\_$DustPol-commander$\_$1024$\_$R2.00.fits} A Venn diagram of the how many SL systems were discovered in each cluster catalog can be found in Figure~\ref{fig:systems-venn-diagram}. Each catalog and their resulting SL systems are described in this section. 

The red sequence Matched-filter Probabilistic Percolation cluster finder algorithm (redMaPPer) \citep{RM2014,mcclintock,y1clustercosmo} identifies galaxy clusters as overdensities of red-sequence galaxies in a catalog. The algorithm counts the excess number of red-sequence galaxies of similar colors that are brighter than a specified luminosity threshold within a circle of radius R$_{\lambda}$ = 1.0 h$^{-1}$ Mpc ($\lambda$/100)$^{0.2}$, where $\lambda$, the ``richness", is the number of galaxies in the cluster \citep{RMrichness}.
The galaxy cluster catalog used in this work was obtained from redMaPPer version 6.4.22+2 on DES Y3A2 Gold data~\citep{y3a2gold}. It is a volume-limited catalog with redshifts of the clusters ranging from $0.04 <  {\rm z}  < 0.95$ and richness $\lambda > 5$. We visually scanned the 21,092 cluster sample with $\lambda>20$ in this search. Though we have previously reported searches of the locations of $\lambda \geq$ 20 redMaPPer clusters in the DES SV and Y1A1 fields~\citep{Nord.Buckley-Geer.ea2016, diehlsvy1}, respectively, we didn't exclude those positions from this search. Of the 152 systems shortlisted for final grading, 143 were ranked $\geq$ 3.
%

The ACT cluster catalog identified more than 4,000 galaxy clusters using the Sunyaev-Zel'dovich effect~\citep{hilton2021}. The full ACT data set~\citep{naess} comprises 18,000 sq. degs. and overlaps the DES Y3 footprint almost completely.  We visually scanned the 1857 ACT galaxy clusters within the DES footprint, identifying 93 for further evaluation and final grading. Of these, 77  were ranked $\geq$ 3. Some of these were already identified in~\cite{hilton2021}.
%

The SPTPol Extended Cluster Survey~\citep{Bleem2020} (SPTPol-ECS) identified  266 galaxy clusters with detection significance $\chi > 5$, and 204 with $4 < \chi < 5$. The full DES footprint partially covered that of SPTPol-ECS. From the 470 total SPTPol-ECS galaxy cluster candidates, we scanned 325 locations in the DES Y6 data. Of these, 17 SL candidate systems were identified for final grading, and all 17 were ranked $\ge 3$.


\section{Search Results}

\begin{figure}[ht]
\begin{center}
{\includegraphics[width=0.32\textwidth]{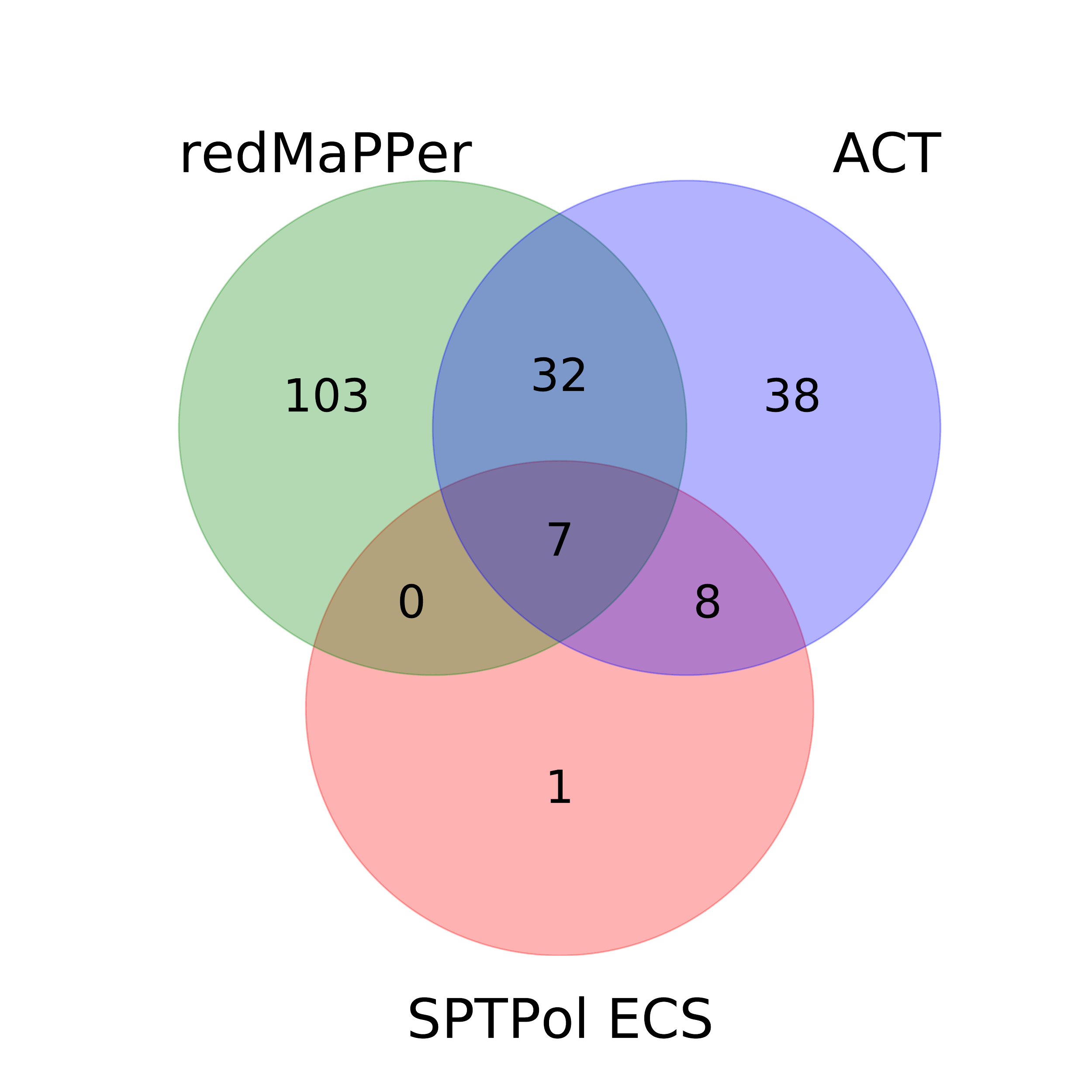}}
{\includegraphics[width=0.32\textwidth]{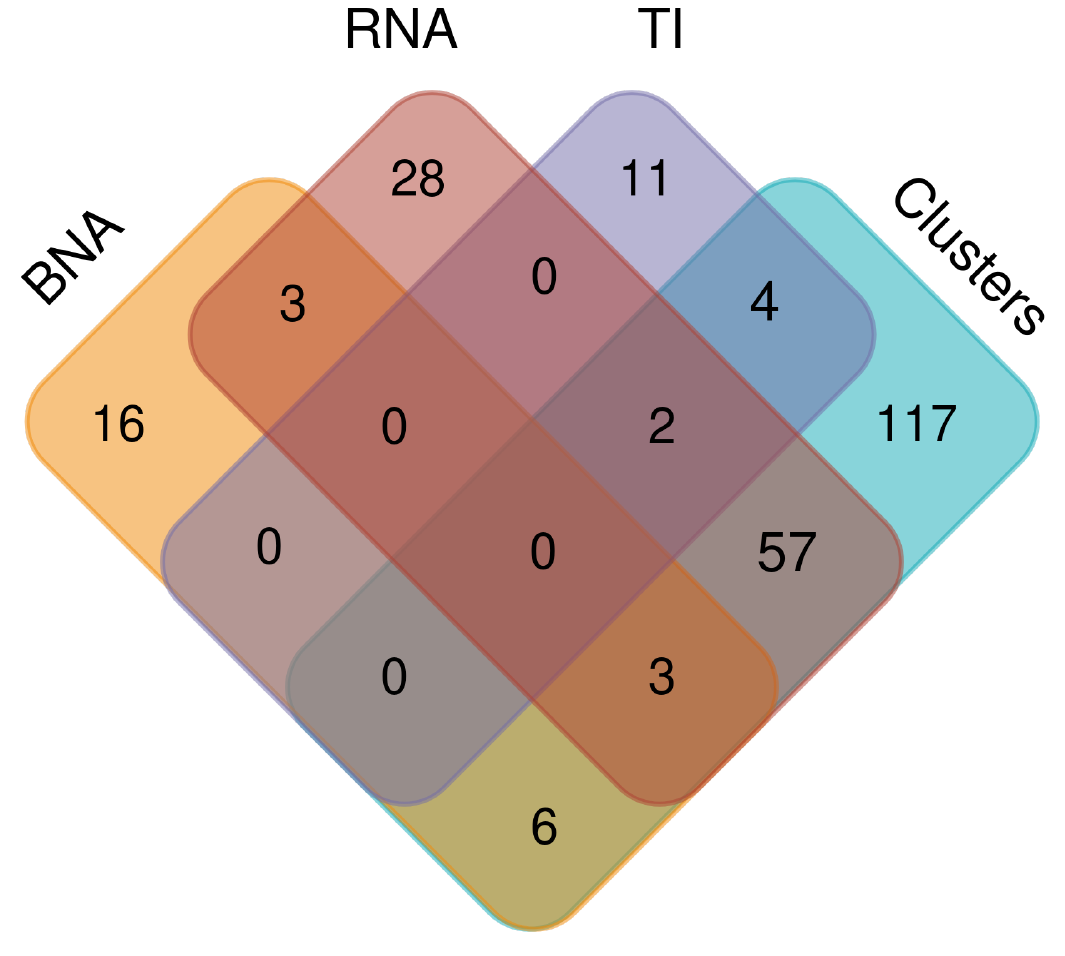}}
{\includegraphics[width=0.32\textwidth]{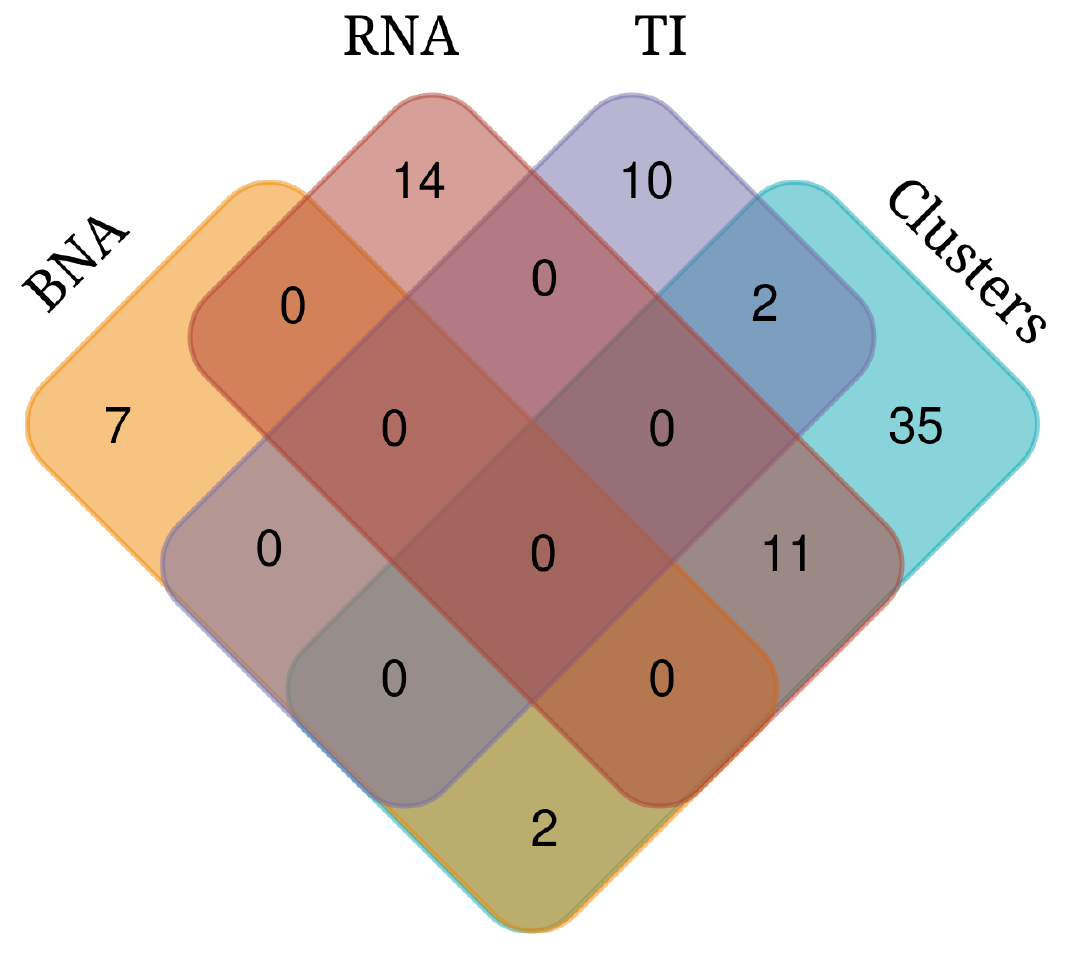}}
\caption{The systems presented in this work, sorted by the algorithms which identified them. \textbf{Left:} a breakdown of the galaxy cluster subset, showing how many systems were identified in redMaPPer, ACT, and SPTPol ECS searches. \textbf{Center:} all 247 systems presented in this work. RNA (BNA) includes both Red  (Blue)-Near-Anything searches, TI are systems found serendipitously in tile inspections, and Clusters includes redMaPPer, ACT, and SPTPol ECS catalogs.
\textbf{Right:} all 81 newly identified systems presented in this work.}
\label{fig:systems-venn-diagram}
\end{center}
\end{figure}

\begin{figure}
\begin{center}
\includegraphics[scale=0.35]{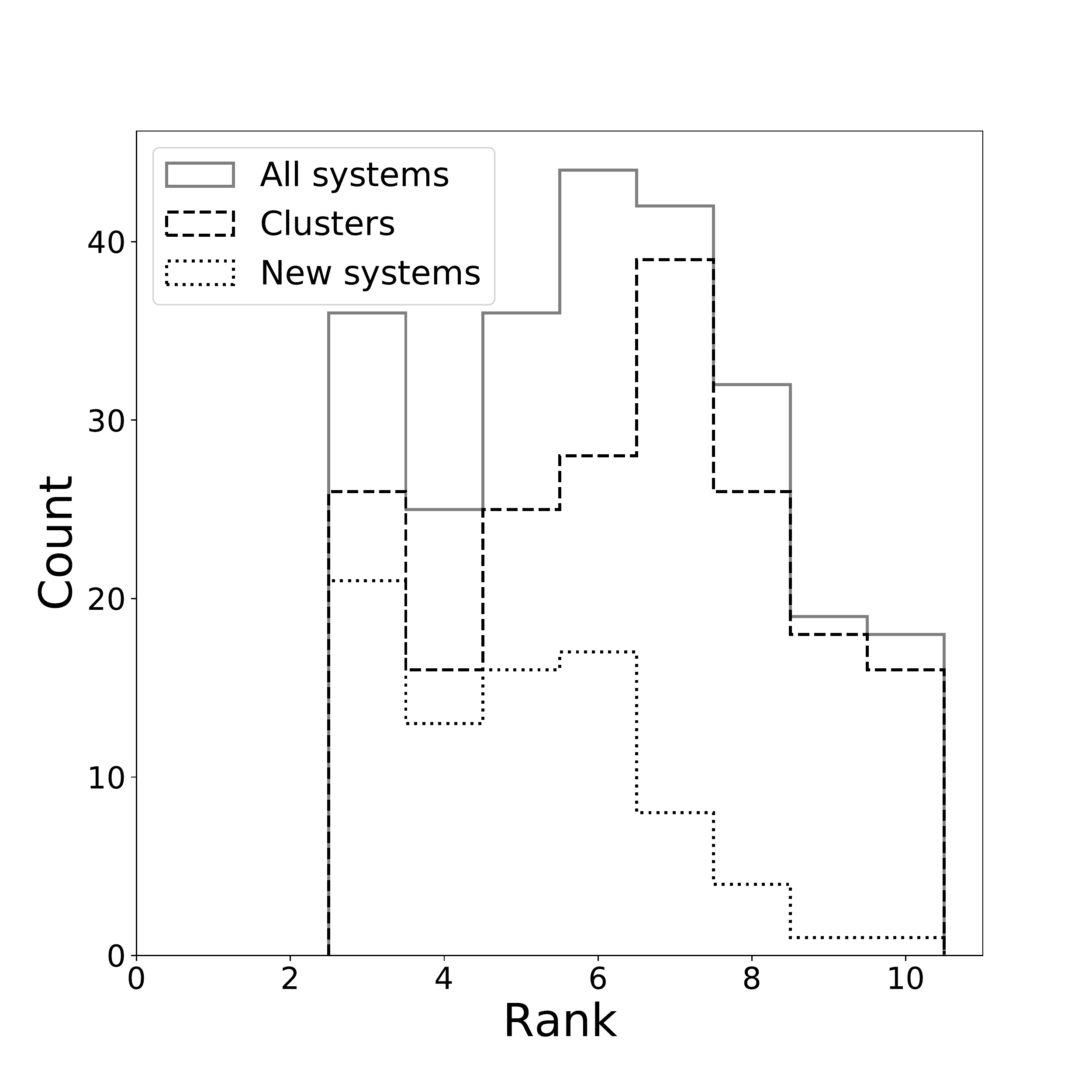}
\caption{The rank of all systems presented in this work. The solid black line shows all systems, the dashed line shows the galaxy clusters subset, and the dotted line shows newly discovered systems. The mean (median) rank of all systems presented is 6.12 (6).}
\label{fig:rank-histogram}
\end{center}
\end{figure}

\begin{figure}[ht]
\begin{center}
\includegraphics[scale=0.35]{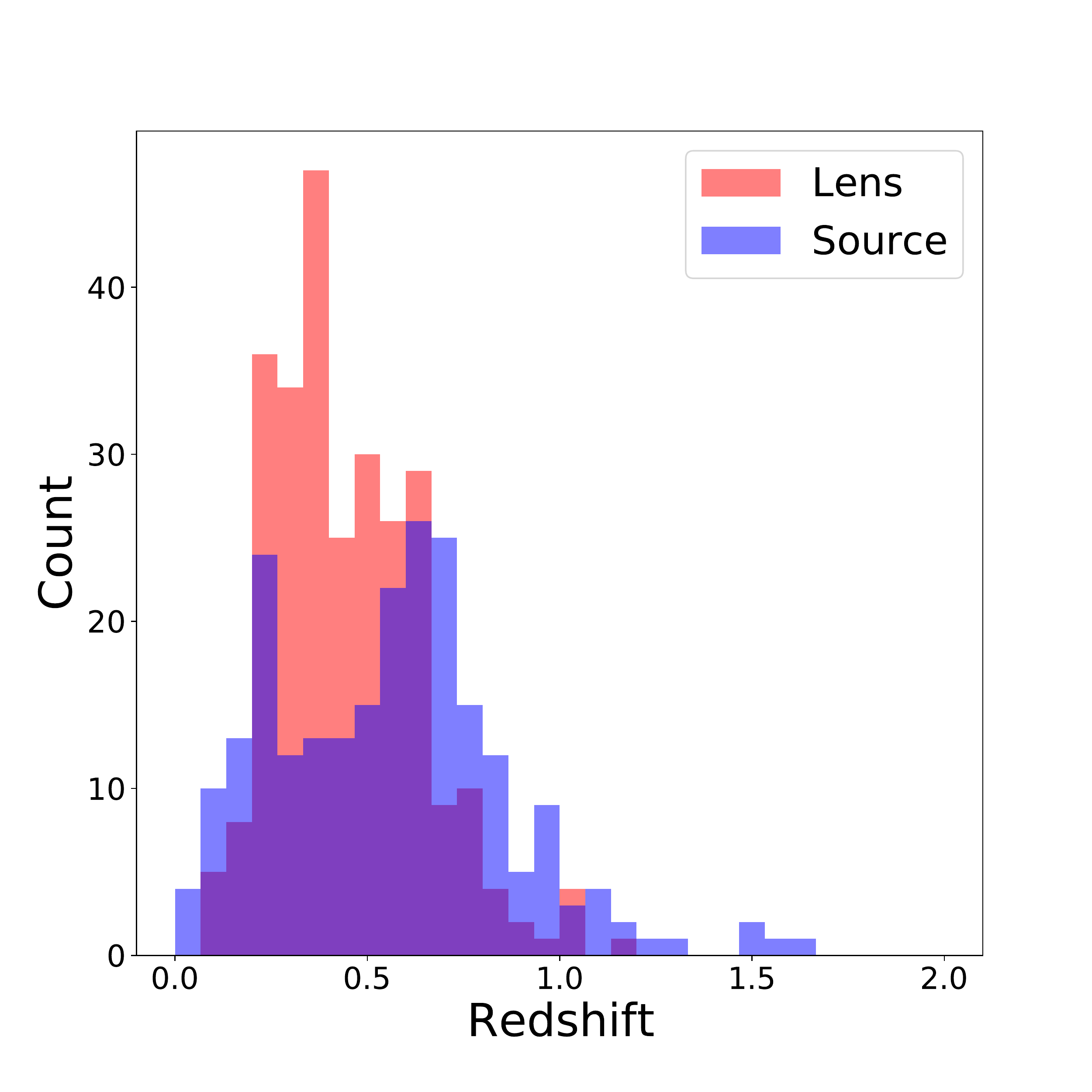}
\caption{DESDM-calculated photometric redshifts for the sources and lenses. As explained in \S 2, the redshifts of the sources are subject to larger uncertainties and systematic errors.  }
\label{fig:photoz}
\end{center}
\end{figure}

\begin{figure}[ht]
\begin{center}
\includegraphics[scale=0.35]{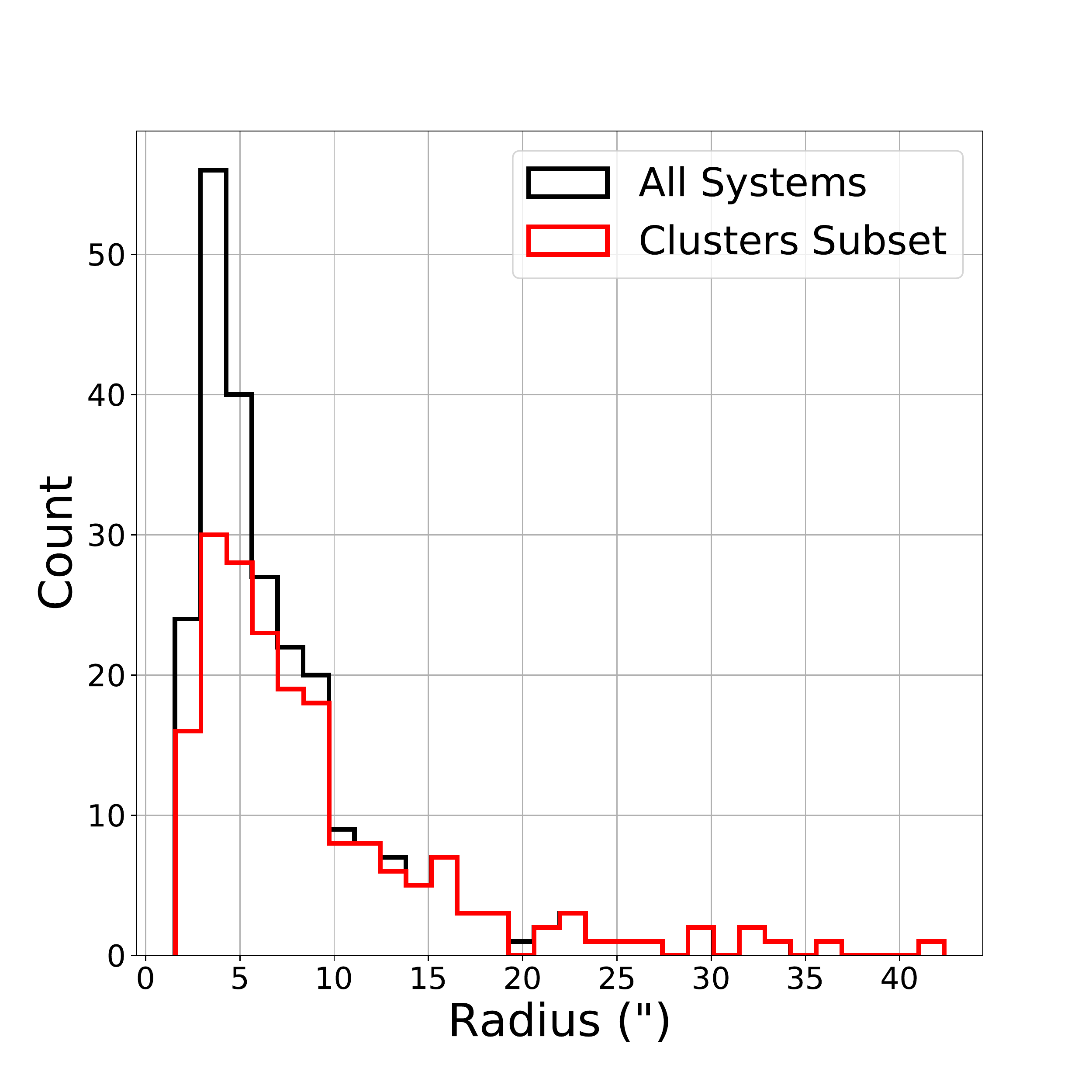}
\caption{The binned distribution of Einstein radius estimates for the lens candidates. The clusters subset, as defined in Figure~\ref{fig:systems-venn-diagram}, is shown separately, and accounts for nearly all systems with an Einstein radius above $10 \arcsec$.}
\label{fig:radius}
\end{center}
\end{figure}

The ranked lists from all searches were combined, yielding 247 SL systems of rank 3 or greater.
Of these systems, 81 are presented for the first time.  Figure~\ref{fig:systems-venn-diagram} shows Venn diagrams illustrating the overlap between each search algorithm or catalog, for all 247 SL candidates and separately for the newly discovered systems.  Table~\ref{tab:numbers} shows the number of candidates searched and systems found for each sub-search.
Figure~\ref{fig:rank-histogram} shows the distribution of the system ranks.
Figures \ref{fig:photoz} and \ref{fig:radius} show the {{lens and source photometric redshifts}}, and Einstein radii for the 247 SL systems. For strong lensing to occur, the source must be strictly more distant than the lens, yet in Figure~\ref{fig:photoz} it is clear that some sources have a lower measured redshift than their respective lenses. For some systems, this discrepancy is likely due to catastrophic redshift errors in blue source objects, as mentioned in Section \ref{sec:des-data}.

\begin{table}[ht]
\begin{center}
\caption{  \label{tab:numbers} Summary of the number of objects visually scanned, the number ranked, and the count of those with rank $\ge3$ for the various searches. We kept track of overlaps between searches for the columns ``\# Rank $ \ge 3 $ and ``\# New", but not for the columns ``\# Scanned" and ``\# Ranked".  Where  there are empty fields, we have not kept track of the distinct counts. Note that a single ``New" system can be discovered by multiple searches. {{Therefore the subtotals and ``Totals'' can be less then the sums of the respective columns.}}   }
\color{black}
\begin{tabular}{|c|ccc|c|} \hline \hline
\textbf{Search}  & \textbf{\# Scanned} & \textbf{\# Ranked} & \textbf{\# Rank $\ge 3$} & \textbf{\# New}  \\ \hline \hline
BNA 3+       & 30319 &  95 & 21  &  6         \\
BBNA 2+      & 18985 &   8 &  7  &  3        \\ \hline
\textbf{BNA Combined} &       & \textbf{103} & \textbf{28}  & \textbf{9}       \\ \hline \hline
RNA 3+ No. 1 & 30994 & 110 & 91   & 25        \\
RNA 3+ No. 2 &  5100 &   4 & 3    & 1        \\ \hline
\textbf{RNA Combined} &       & \textbf{114} & \textbf{93}  & \textbf{25}        \\ \hline \hline
\textbf{Tile Inspection}
             &  \textbf{ 4085} &  \textbf{44} &  \textbf{17} & \textbf{12}        \\ \hline \hline
RedMapper    & 21092 & 152 & \textbf{142} & \textbf{35}      \\
ACT          &  1857 &  93 &  \textbf{85} & \textbf{21}        \\
SPTPOL       &  325  &  17 &  \textbf{16} & \textbf{2}        \\ \hline
\textbf{Clusters Combined}
             &       & \textbf{202} &  \textbf{189}  & \textbf{50}        \\ \hline \hline
\textbf{Total}        &       &     &  \textbf{247}  & \textbf{81}     \\ \hline \hline
\end{tabular}
\end{center}
\end{table}

\color{black}

\color{red}

\color{black}{}

\begin{figure}[ht]
\begin{center}
\includegraphics[scale=0.26]{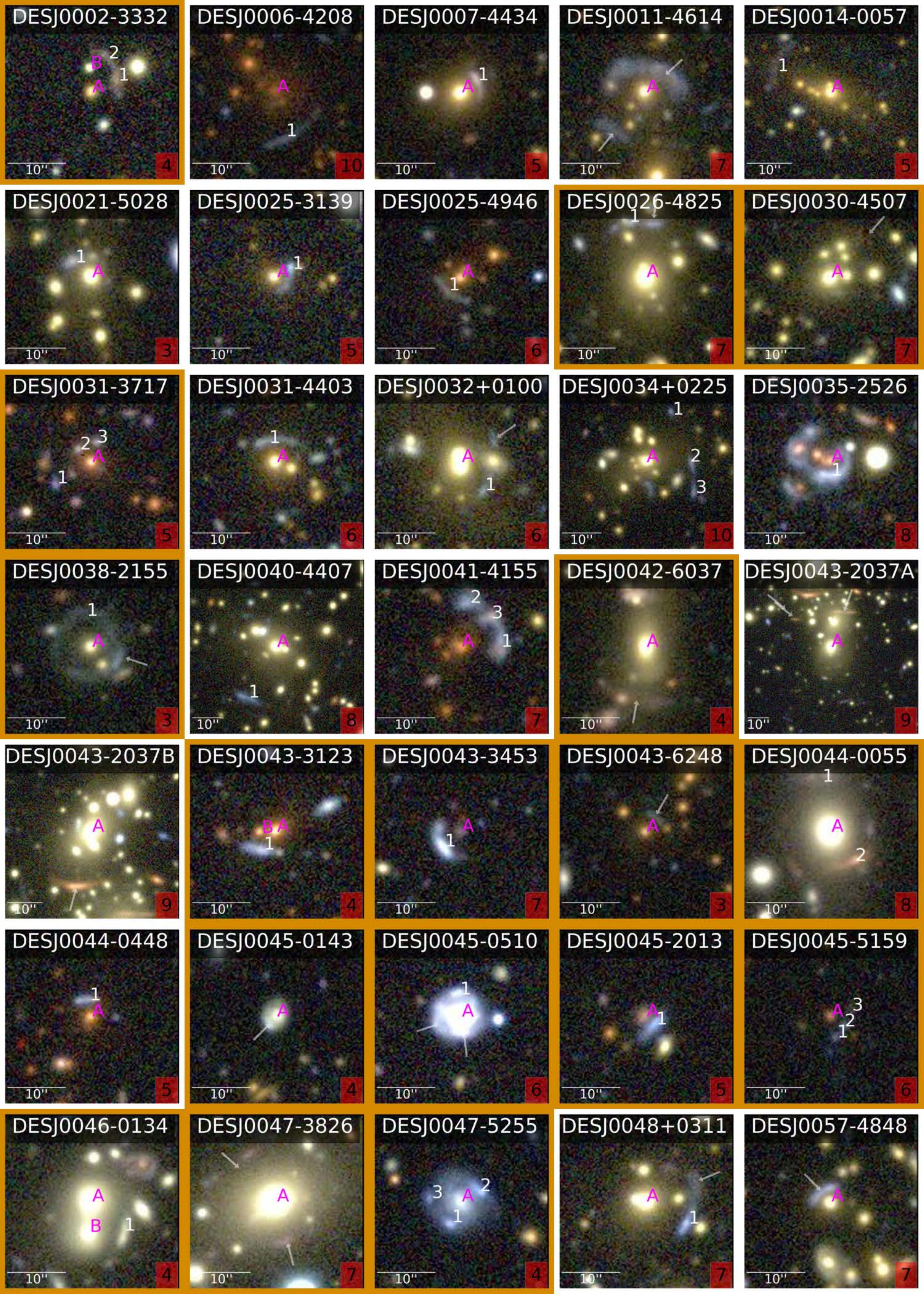}
\caption{First page of SL systems with rank 3 or more.   Each cutout image has the visual inspection ranking displayed in a red box in the lower right hand corner. All images are oriented north up, east left.  Each cutout is dynamically sized to fit the SL system well within the cutout, with a minimum size of 30\arcsec \ $\times$ 30\arcsec.  A scale bar 10\arcsec \ long is displayed in the lower left hand corner. 
\bf{New SL systems are outlined with a gold border, previously known systems with white.}}
\label{fig:cutouts-1}
\end{center}
\end{figure}

\begin{figure}[ht]
\begin{center}
\includegraphics[scale=0.26]{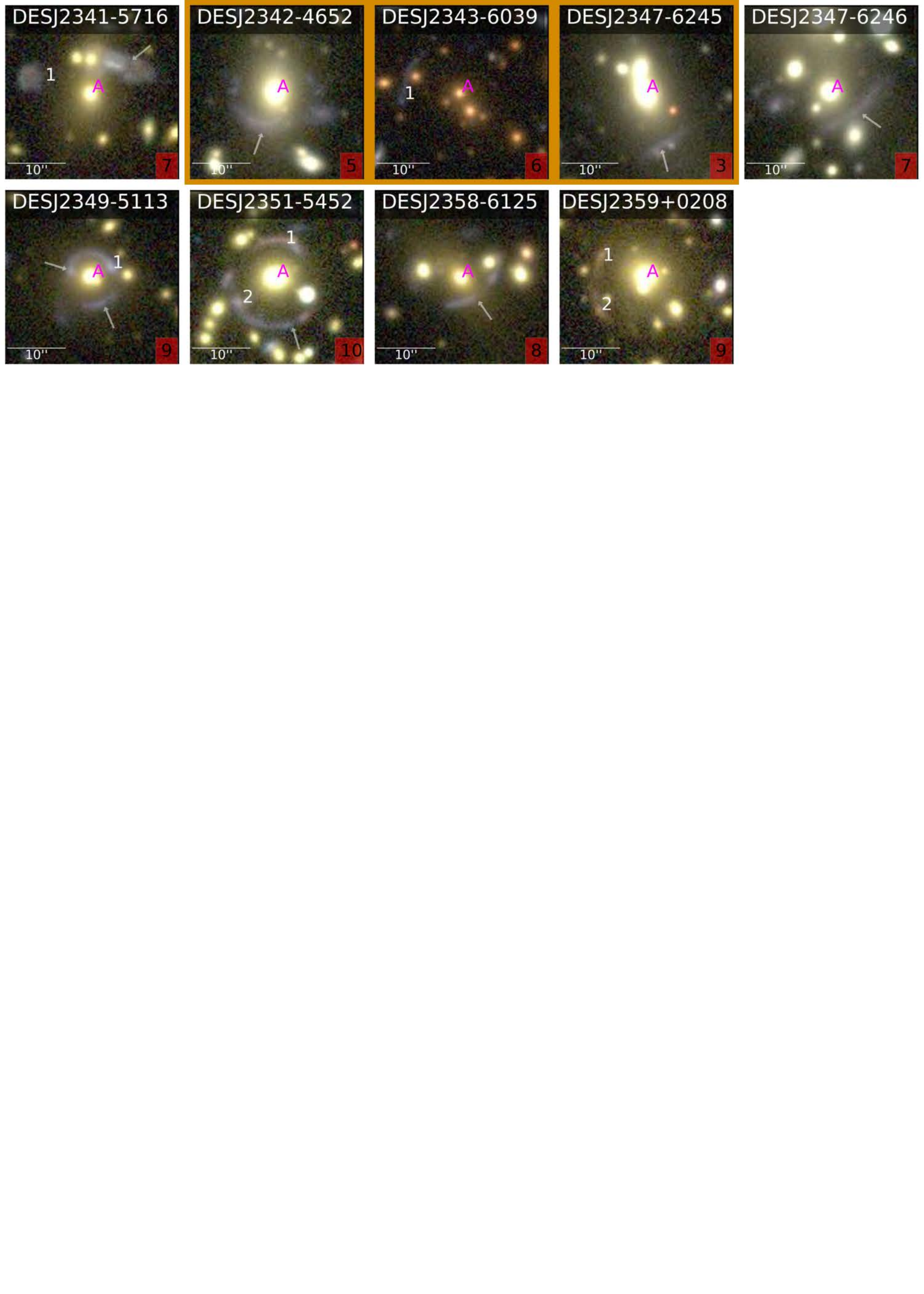}
\caption{Eighth and last page of SL systems with rank 3 or more.   Each cutout image has the visual inspection ranking displayed in a red box in the lower right hand corner. All images are oriented north up, east left.  Each cutout is dynamically sized to fit the SL system well within the cutout, with a minimum size of 30\arcsec \ $\times$ 30\arcsec.  A scale bar 10\arcsec \ long is displayed in the lower left hand corner. New SL systems are outlined with a gold border, previously known systems with white. }
\label{fig:cutouts-8}
\end{center}
\end{figure}

Table \ref{tab:all_search_systems} provides an overview of each candidate lens system, including: the system name, algorithms or catalogs which identified the candidates for scanning, system rank, Einstein radius estimate (described below), and references to previous discoveries of the system. Figures \ref{fig:cutouts-1}  through \ref{fig:cutouts-8} show a 3-color cutout of each system. The cutouts are made from the DES Y6A1 coadded tiles using the i, r, and g-bands for the RGB color channels respectively. Where SL systems fell across tile edges, multiple coadd tiles were stitched together to form one cutout. Each cutout was dynamically scaled to fit the system well within the cutout based on the radial distance from the lens center to the furthest
{{lensed source image.}}

Each image is centered on the most likely lens galaxy, labeled by the letter ``A". For systems where the primary lens was uncertain, or where sources were clearly perturbed by multiple galaxies, additional lens galaxies are marked with roman letters ``B", ``C", etc. All lenses are found in the DESDM Y3 or Y6 catalogs. Source objects which were identified in DESDM are labeled by number, e.g. ``1", ``2", etc. Source objects which were not identified in DESDM had their positions determined manually, and are shown by an arrow.

To identify positions and photometry for lens and source objects, the DES Y3 and Y6 catalogs were overlaid on images of candidates in the NCSA target viewer. Lens and source images were matched to catalog object IDs. Y6 object identifications were preferred over Y3; however, in cases where an object was not detected in Y6, or was blended with another object, the object was matched to the Y3 catalog. Within each system, catalogs were kept consistent, such that no system mixes Y3 data for some objects with Y6 data for others. Some source images could not be accurately associated with a known DES object, and their position was estimated manually. Table \ref{tab:all_search_objects} shows the positions and photometry for all systems presented in this paper.


We identified those systems presented here that were previously known, as well as those that are new. To do so, we checked the candidates against the Master Lens Database~\citep{MLD}, the list of objects labeled as strong lens candidates or known systems in SIMBAD \citep{simbad}, and other SL search papers. Previous discoveries of all systems are given in Table~\ref{tab:all_search_systems}. 
Note that references include both confirmed and candidate strong lens systems, i.e. systems with external references are not guaranteed to be true lenses. We include these previously discovered systems in this work because we provide positions, photometry, and photometric redshifts of system components (see Table~\ref{tab:all_search_objects}), as well as an estimate of the Einstein radius (see Table~\ref{tab:all_search_systems}), forming a standardized dataset which may be of use in further studies.

The average radius of source images is an approximation for the Einstein radius, and is identical to that when the true source position is directly behind
the lens. We estimate the Einstein radius from the mean separation between the primary lens and each source image.  For systems with multiple source images, the uncertainty is taken as the standard deviation of these separations. For systems with only one source image, the uncertainty is taken as 10\% of the source-lens separation,
{{typical for the scatter of individual sources in systems with multiple images. Especially for these systems with a single lensed image, this tends to provide an overestimate of the Einstein radius.}~\citep{KneibNatarajan2011}.}
In all cases, the uncertainty is summed in quadrature with the DES plate scale of 0.263\arcsec \ per pixel.
The image separation distribution is sensitive to a number of inputs such as the halo mass, the lens mass distribution, and the source redshift. It therefore contains information about the cosmological parameters and various scaling relations between galaxy properties and halo mass, and can be measured from galaxy to cluster scales~\citep{Oguri2006, sw2-2016}. A similar metric has been found to be an effective measure for the inner halo mass by \citet{RemolinaGonzalez2020}.

\startlongtable
\begin{deluxetable*}{lclCc}
    \tablecaption{Properties of All Candidate Lensing Systems, ordered by increasing RA. \label{tab:all_search_systems} }
    \tabletypesize{\scriptsize}
    \tablehead{
    \colhead{System Name} & \colhead{Rank} & \colhead{Algorithm} & \colhead{$\theta_E$ (\arcsec)} & \colhead{References}
    }

    \startdata
        DESJ0002-3332 & 4 & BNA & 5.69 \pm 0.83 & \ldots \\
DESJ0006-4208 & 10 & ACT & 7.70 \pm 0.81 & (B) \\
DESJ0007-4434 & 5 & RedM & 3.43 \pm 0.43 & (B), (D) \\
DESJ0011-4614 & 7 & ACT, RedM & 6.53 \pm 2.35 & (B), (D) \\
DESJ0014-0057 & 5 & ACT, BNA, RedM missed & 12.84 \pm 1.31 & SOGRAS J0014-0057 (h), (w) \\
DESJ0021-5028 & 3 & RNA, RedM missed & 3.87 \pm 0.47 & (B) \\
DESJ0025-3139 & 5 & RedM & 2.98 \pm 0.40 & (D) \\
DESJ0025-4946 & 6 & RNA2 & 3.21 \pm 0.41 & DES J002510-494626 (A) \\
DESJ0026-4825 & 7 & ACT, RedM & 10.89 \pm 0.26 & \ldots \\
DESJ0030-4507 & 7 & RedM & 9.71 \pm 1.01 & \ldots \\
DESJ0031-4403 & 6 & RNA & 3.56 \pm 0.44 & DES J003104-440300 (A), (B) \\
DESJ0031-3717 & 5 & RNA & 4.53 \pm 1.85 & \ldots \\
DESJ0032+0100 & 6 & RNA, ACT, RedM & 6.72 \pm 0.51 & HSCJ003217+010037 (w), (z) \\
DESJ0034+0225 & 10 & RedM & 13.22 \pm 1.59 & HSCJ003428+022522 (w), (z) \\
DESJ0035-2526 & 8 & BNA & 2.34 \pm 0.35 & (D) \\
DESJ0038-2155 & 3 & BNA & 5.64 \pm 0.33 & \ldots \\
DESJ0040-4407 & 8 & ACT, RedM missed & 18.67 \pm 1.89 & (B) \\
DESJ0041-4155 & 7 & BNA, ACT & 7.23 \pm 0.47 & (B) \\
DESJ0042-6037 & 4 & TI, RedM missed & 9.19 \pm 0.96 & \ldots \\
DESJ0043-6248 & 3 & TI & 3.22 \pm 0.42 & \ldots \\
DESJ0043-2037A & 9 & TI, ACT, SPTPol & 22.71 \pm 4.16 & SPT-CL J0043-2037 (m), (z) \\
DESJ0043-2037B & 9 & TI, ACT, SPTPol & 16.69 \pm 1.69 & SPT-CL J0043-2037 (m), (z) \\
DESJ0043-3123 & 4 & TI & 3.85 \pm 0.47 & \ldots \\
DESJ0043-3453 & 7 & TI & 3.90 \pm 0.47 & \ldots \\
DESJ0044-0448 & 5 & TI & 2.86 \pm 0.39 & DESI-011.0219-04.8058 (v) \\
DESJ0044-0055 & 8 & TI, RedM & 7.63 \pm 1.18 & \ldots \\
DESJ0045-0510 & 6 & TI & 3.58 \pm 0.57 & \ldots \\
DESJ0045-2013 & 5 & TI & 2.09 \pm 0.34 & \ldots \\
DESJ0045-0143 & 4 & TI & 1.56 \pm 0.31 & \ldots \\
DESJ0045-5159 & 6 & TI & 3.43 \pm 0.49 & \ldots \\
DESJ0046-0134 & 4 & TI, RedM missed & 7.40 \pm 0.79 & \ldots \\
DESJ0047-5255 & 4 & TI & 4.26 \pm 0.73 & \ldots \\
DESJ0047-3826 & 7 & TI, ACT, RedM missed & 7.88 \pm 0.72 & \ldots \\
DESJ0048+0311 & 7 & TI, RedM, RNA2 & 8.80 \pm 0.72 & RCSGA J004827+031114 (g), (D) \\
DESJ0057-4848 & 7 & RNA, RedM & 2.47 \pm 0.36 & (B) \\
DESJ0059-2800 & 4 & RedM & 7.28 \pm 0.77 & \ldots \\
DESJ0100-3936 & 3 & RedM & 6.92 \pm 1.07 & \ldots \\
DESJ0102-2911 & 8 & RedM & 4.89 \pm 0.56 & KiDS J010257.486-291121.76 (x), (z), (D) \\
DESJ0103-2234 & 3 & RedM & 9.17 \pm 0.95 & \ldots \\
DESJ0103-1821 & 7 & RedM & 9.33 \pm 0.97 & DESI-015.8440-18.3629 (z) \\
DESJ0104-5341 & 7 & RedM & 2.53 \pm 0.37 & (B) \\
DESJ0105-3939 & 4 & RedM & 3.16 \pm 0.41 & (D) \\
DESJ0105-3520 & 3 & RNA & 2.24 \pm 0.69 & \ldots \\
DESJ0106-3104 & 7 & RedM & 2.97 \pm 0.40 & KiDS J010606.232-310437.84 (x), (z) \\
DESJ0106-3700 & 3 & ACT, RedM & 5.92 \pm 0.65 & DESI-016.5369-37.0055 (z) \\
DESJ0106-5355 & 10 & RNA, ACT, RedM & 11.24 \pm 2.08 & (B) \\
DESJ0107-3128 & 10 & BNA, RNA & 5.04 \pm 0.61 & KiDS J010704.918-312841.03 (x), (z) \\
DESJ0109-3335 & 9 & RedM & 8.59 \pm 0.90 & (D) \\
DESJ0115-3520 & 3 & RNA & 4.99 \pm 0.56 & (D) \\
DESJ0120-5143 & 10 & BNA & 3.35 \pm 0.49 & DES J012042-514353 (A), (B), (D) \\
DESJ0128-2905 & 8 & RedM & 5.23 \pm 0.58 & LinKS 2323 (u) \\
DESJ0128-2957 & 5 & RNA, RedM & 3.67 \pm 0.57 & DESI-022.2123-29.9602 (z) \\
DESJ0129-1641 & 4 & RNA, ACT & 3.15 \pm 0.41 & \ldots \\
DESJ0130-3744 & 7 & BNA, RNA, RedM & 5.33 \pm 0.91 & DES J013002-374457 (A), (D) \\
DESJ0131-1336 & 3 & ACT, BBNA & 22.78 \pm 2.29 & CLASH A209 (l) \\
DESJ0133-1650 & 3 & BNA & 5.70 \pm 0.63 & DESI-023.4239-16.8390 (z) \\
DESJ0133-6434 & 8 & RNA, BBNA & 4.44 \pm 0.46 & (D) \\
DESJ0134+0433 & 3 & RedM & 4.87 \pm 0.55 & DESI-023.6765+04.5639 (v) \\
DESJ0135-2328 & 4 & ACT & 6.39 \pm 0.69 & \ldots \\
DESJ0135-2033 & 7 & RedM & 4.75 \pm 0.60 & DES J013542-203335 (A), (D) \\
DESJ0137-1034 & 4 & BNA, RNA, ACT, RedM missed & 13.95 \pm 1.42 & DESI-024.2940-10.5728 (z), (A) \\
DESJ0138-2155 & 10 & ACT, SPTPol & 18.18 \pm 0.95 & SPT-CL J0138-2155 (m) \\
DESJ0138-2844 & 8 & RedM & 3.54 \pm 0.44 & DESI-024.5974-28.7358 (z), (A), (D) \\
DESJ0142-5032 & 9 & BNA, ACT & 14.03 \pm 0.64 & (B) \\
DESJ0143-0850 & 6 & RNA & 2.47 \pm 0.37 & DES J014326-085021 (A), (C), (D) \\
DESJ0143-2617 & 4 & RedM & 5.80 \pm 0.64 & DESI-025.9390-26.2946 (z) \\
DESJ0144-2213 & 7 & RNA, ACT, RedM, SPTPol & 11.88 \pm 1.22 & SPT-CL J0144-2214 (m) \\
DESJ0145-0455 & 5 & RNA & 2.34 \pm 0.35 & CSWA 103 (i), (v), (y), (D) \\
DESJ0145+0137 & 3 & RedM & 10.13 \pm 1.05 & \ldots \\
DESJ0145-3541 & 10 & RedM & 3.27 \pm 0.42 & DES J014546-354127 (A), (D) \\
DESJ0146-0929 & 10 & ACT, RedM & 12.05 \pm 1.17 & Hall's Arc (b), (i), (D) \\
DESJ0149-3825 & 6 & ACT, SPTPol & 9.42 \pm 0.98 & (D) \\
DESJ0151-3544A & 8 & RNA, ACT, RedM, SPTPol & 7.32 \pm 0.78 & SPT-CL J0151-3544 (m) \\
DESJ0151-3544B & 8 & RNA, ACT, RedM, SPTPol & 29.34 \pm 2.95 & SPT-CL J0151-3544 (m) \\
DESJ0152-2853 & 3 & ACT, SPTPol & 16.42 \pm 0.57 & \ldots \\
DESJ0154-2324 & 7 & RedM & 3.59 \pm 0.30 & DESI-028.6096-23.4067 (z) \\
DESJ0158-0040 & 6 & RedM & 3.52 \pm 0.84 & DESI-029.6032-00.6665 (v), (D) \\
DESJ0159-3413 & 10 & ACT, RedM, SPTPol & 10.39 \pm 0.63 & SPT-CL J0159-3413 (m), (D) \\
DESJ0200+0127 & 6 & RNA, RedM missed & 3.99 \pm 0.48 & \ldots \\
DESJ0203-2338 & 7 & BNA & 3.86 \pm 0.45 & PS1J0203-2338 (y), (A), (C), (D) \\
DESJ0203-2017 & 7 & RNA, ACT, RedM missed & 5.95 \pm 0.65 & SPT-CL J0203-2017 (m) \\
DESJ0203-3104 & 6 & RNA, RedM & 2.37 \pm 0.38 & DESI-030.9153-31.0823 (z) \\
DESJ0205-3539 & 8 & RedM & 3.36 \pm 0.43 & (D) \\
DESJ0205-1935 & 6 & RNA, ACT, RedM & 4.30 \pm 0.50 & \ldots \\
DESJ0206+0258 & 6 & RNA, RedM & 4.14 \pm 0.49 & \ldots \\
DESJ0206-0114 & 5 & RNA, ACT & 2.22 \pm 0.45 & HSCJ020613-011417 (w), (D) \\
DESJ0207-2726 & 8 & RedM & 4.34 \pm 0.69 & DESI-031.7778-27.4457 (z), (D) \\
DESJ0209+0222 & 4 & ACT, RedM & 7.73 \pm 0.82 & HSCJ020937+022256 (w) \\
DESJ0209-3547 & 6 & ACT, RedM & 5.92 \pm 0.65 & DESI-032.4765-35.7990 (z) \\
DESJ0212-2842 & 3 & RedM & 6.47 \pm 0.57 & \ldots \\
DESJ0214-0206 & 3 & RNA & 3.14 \pm 0.43 & HSCJ021408-020628 (w), (y), (D) \\
DESJ0214-0535 & 7 & RedM & 7.30 \pm 0.82 & SL2S J021408-053532 (a), (w) \\
DESJ0214-4207 & 4 & RedM & 2.45 \pm 0.36 & \ldots \\
DESJ0218-3142 & 6 & RNA, ACT, SPTPol, RedM missed & 8.50 \pm 0.89 & \ldots \\
DESJ0219-4427 & 6 & RedM & 3.87 \pm 0.47 & \ldots \\
DESJ0219+0247 & 3 & ACT, RedM missed & 17.12 \pm 1.73 & HSCJ021953+024707 (w) \\
DESJ0220-3833 & 8 & RedM & 8.98 \pm 2.14 & DESI-035.2405-38.5511 (z), (D) \\
DESJ0225-4200 & 7 & RNA & 10.68 \pm 1.10 & \ldots \\
DESJ0227-4516 & 7 & RedM & 3.99 \pm 0.48 & (B) \\
DESJ0229-3110 & 8 & RedM & 5.21 \pm 0.58 & KiDS J022956.259-311022.65 (x), (z), (A) \\
DESJ0230-2702 & 4 & RNA, ACT, RedM missed & 6.42 \pm 0.69 & \ldots \\
DESJ0234-4529 & 3 & BNA & 3.64 \pm 0.73 & (B) \\
DESJ0237-3017 & 5 & RNA, RedM & 5.05 \pm 0.57 & DESI-039.3484-30.2921 (z) \\
DESJ0238-3334 & 5 & BNA & 3.73 \pm 0.26 & \ldots \\
DESJ0239-0127 & 3 & RedM & 5.40 \pm 0.92 & DESI-039.9261-01.4632 (v), (w) \\
DESJ0239-0134 & 10 & RNA, ACT, RedM & 11.47 \pm 1.12 & Abell 370 (t), (w), (D) \\
DESJ0242-2132 & 6 & RNA & 19.33 \pm 0.47 & \ldots \\
DESJ0243-3843 & 7 & RNA, ACT, RedM missed & 15.20 \pm 1.19 & SPT-CL J0243-4833 (p) \\
DESJ0245-5301 & 6 & ACT, RedM & 6.21 \pm 1.90 & (D) \\
DESJ0248-0331 & 10 & ACT & 15.92 \pm 0.46 & CLASH A383 (l), (D) \\
DESJ0248-0216 & 10 & ACT, RedM & 14.21 \pm 0.81 & DESI-042.0371-02.2771 (z) \\
DESJ0248-3955 & 6 & BNA & 3.95 \pm 0.40 & (D) \\
DESJ0250-6308 & 8 & RNA & 4.19 \pm 0.49 & \ldots \\
DESJ0251-5515 & 6 & RedM & 6.89 \pm 0.74 & (B) \\
DESJ0252-4732 & 7 & RedM & 2.94 \pm 0.39 & (B), (D) \\
DESJ0252-1459 & 6 & RedM & 3.44 \pm 0.43 & \ldots \\
DESJ0253-2629 & 5 & RedM & 4.66 \pm 0.54 & \ldots \\
DESJ0257-2209 & 5 & RNA, ACT, SPTPol & 1.83 \pm 0.32 & SPT-CL J0257-2209 (m), (z) \\
DESJ0304-4921 & 10 & ACT, RedM & 25.42 \pm 0.95 & (B), (D) \\
DESJ0307-5042 & 5 & ACT, RedM missed & 11.57 \pm 1.00 & (B), (D) \\
DESJ0309-3805 & 5 & RedM & 1.71 \pm 0.31 & DES J030920-380545 (A), (D) \\
DESJ0310-4647 & 8 & RNA, ACT & 9.49 \pm 0.37 & SPT-CL J0310-4647 (g), (B), (D) \\
DESJ0311-4232 & 6 & RNA, RedM missed & 5.13 \pm 0.58 & (D) \\
DESJ0315-5954 & 8 & RedM & 8.84 \pm 0.92 & \ldots \\
DESJ0316-4816 & 4 & BNA & 3.01 \pm 0.48 & (B) \\
DESJ0317-4421 & 3 & ACT & 7.81 \pm 0.82 & \ldots \\
DESJ0327-1326 & 10 & RNA, ACT, RedM & 16.01 \pm 2.69 & RCSGA 032727-132609 (d), (D) \\
DESJ0328-2140 & 6 & RNA, ACT, RedM, SPTPol & 21.14 \pm 3.23 & (D) \\
DESJ0330-5228 & 9 & TI, ACT, RedM & 6.21 \pm 0.27 & DES J0330-5228 (o), (B), (D) \\
DESJ0334-1311 & 6 & RedM & 6.84 \pm 1.16 & DESI-053.6254-13.1866 (z) \\
DESJ0336-3812 & 3 & RedM & 1.58 \pm 0.31 & (D) \\
DESJ0339-4849 & 5 & ACT, RedM & 12.65 \pm 1.10 & (B) \\
DESJ0339-3856 & 5 & RNA, RedM & 1.85 \pm 0.32 & DESI-054.8226-38.9498 (z) \\
DESJ0339-3313 & 5 & RedM & 11.85 \pm 1.21 & \ldots \\
DESJ0341-5130 & 8 & RedM & 2.58 \pm 0.37 & (B), (D) \\
DESJ0342-5355A & 8 & ACT, RedM & 4.47 \pm 0.52 & (B), (D) \\
DESJ0342-5355B & 8 & ACT, RedM & 11.24 \pm 0.46 & (B), (D) \\
DESJ0342-2342 & 5 & RNA & 5.51 \pm 0.72 & \ldots \\
DESJ0345-3403 & 4 & ACT, RedM & 9.82 \pm 3.65 & \ldots \\
DESJ0348-2145 & 9 & RNA, ACT, RedM, SPTPol & 8.72 \pm 0.83 & SPT-CL J0348-2144 (m), (D) \\
DESJ0352-3825 & 8 & RedM & 3.18 \pm 0.43 & DES J035242-382544 (A), (D) \\
DESJ0354-4446 & 7 & RNA, TI, RedM & 3.81 \pm 0.46 & (B) \\
DESJ0356-5607 & 5 & RedM & 3.92 \pm 0.44 & (D) \\
DESJ0356-2408 & 7 & RedM & 4.97 \pm 2.40 & DES J035649-240841 (A), (D) \\
DESJ0357-4756 & 10 & RedM & 8.63 \pm 1.30 & (B) \\
DESJ0358-2415 & 3 & RNA, RNA2, ACT & 13.05 \pm 1.21 & \ldots \\
DESJ0359-2433 & 5 & RNA, ACT, RedM & 10.29 \pm 1.06 & \ldots \\
DESJ0402-2205 & 9 & RNA, RedM & 5.10 \pm 0.57 & DESI-060.5238-22.0990 (z), (A), (D) \\
DESJ0403-5057 & 6 & RNA, RedM & 4.16 \pm 0.47 & (B) \\
DESJ0406-2254 & 6 & RNA & 4.94 \pm 0.60 & DESI-061.5534-22.9034 (z) \\
DESJ0411-4506 & 3 & RNA, RedM & 4.21 \pm 0.50 & (B) \\
DESJ0411-4819 & 9 & RNA, ACT, RedM & 7.08 \pm 0.54 & SPT-CL J0411-4819 (p), (B), (D) \\
DESJ0413-1958 & 6 & RNA & 3.32 \pm 0.33 & \ldots \\
DESJ0413-5117 & 7 & ACT & 2.79 \pm 0.38 & \ldots \\
DESJ0415-4143 & 7 & ACT & 6.72 \pm 0.27 & (C), (D) \\
DESJ0416-2404 & 3 & RNA, ACT, SPTPol, RedM missed & 5.58 \pm 0.62 & MACS J0416.1-2403 (j) \\
DESJ0416-5525 & 7 & RNA, RedM & 4.34 \pm 0.51 & (B), (D) \\
DESJ0422-2803 & 8 & RNA & 2.53 \pm 0.36 & DESI-065.6447-28.0652 (z), (A) \\
DESJ0424-3317 & 9 & RNA, RedM, BBNA & 5.49 \pm 0.29 & (D) \\
DESJ0438-3220 & 4 & RNA, ACT, RedM & 4.84 \pm 0.55 & DESI-069.5641-32.3475 (z) \\
DESJ0440-2658 & 6 & RNA, RedM & 10.55 \pm 1.37 & DESI-070.1928-26.9754 (z) \\
DESJ0440-4657 & 6 & RNA, ACT, RedM & 8.36 \pm 0.88 & (B) \\
DESJ0448-3019 & 6 & ACT, RedM, SPTPol & 13.98 \pm 0.41 & DESI-072.0528-30.3308 (z) \\
DESJ0451-1856 & 3 & RNA, ACT, RedM & 21.89 \pm 2.20 & \ldots \\
DESJ0453-4120 & 3 & RedM & 2.22 \pm 0.34 & \ldots \\
DESJ0453-3639 & 3 & RedM & 3.26 \pm 0.42 & \ldots \\
DESJ0455-2530 & 8 & RNA, RedM & 16.07 \pm 0.49 & DESI-073.9030-25.5129 (z), (D) \\
DESJ0459-3756 & 5 & RNA, RedM missed & 9.18 \pm 0.96 & \ldots \\
DESJ0501-2425 & 8 & RNA, RedM & 27.39 \pm 2.75 & DESI-075.2790-24.4179 (z) \\
DESJ0509-5342 & 6 & ACT, RedM & 9.52 \pm 0.88 & (B), (D) \\
DESJ0510-3232 & 7 & RedM & 3.90 \pm 0.33 & (D) \\
DESJ0512-3847 & 8 & RNA, SPTPol, RedM missed & 28.82 \pm 4.04 & SPT-CL J0512-3848 (m), (D) \\
DESJ0513-2128 & 5 & RNA, RedM missed & 5.39 \pm 0.60 & DESI-078.3580-21.4717 (z), (D) \\
DESJ0513-3050 & 9 & RedM & 6.11 \pm 0.54 & DESI-078.3561-30.8433 (z) \\
DESJ0514-5142 & 5 & RNA, RedM & 5.26 \pm 0.59 & (B) \\
DESJ0516-2208 & 4 & RNA, RedM & 2.39 \pm 0.36 & DES J051603-220847 (A), (D) \\
DESJ0517-2526 & 5 & RedM & 5.77 \pm 0.63 & \ldots \\
DESJ0518-4348 & 6 & BNA & 3.76 \pm 0.46 & \ldots \\
DESJ0524-2721 & 6 & RedM & 6.56 \pm 0.47 & (D) \\
DESJ0525-4424 & 8 & RNA & 3.38 \pm 0.43 & (B), (D) \\
DESJ0525-3712 & 5 & RedM & 5.56 \pm 0.61 & \ldots \\
DESJ0527-1858 & 7 & RedM & 6.79 \pm 1.45 & DESI-081.7547-18.9674 (z) \\
DESJ0528-2633 & 6 & RNA & 4.69 \pm 0.89 & DESI-082.1548-26.5667 (z) \\
DESJ0528-3958 & 6 & RNA & 5.31 \pm 0.93 & \ldots \\
DESJ0531-3158 & 5 & RedM & 5.33 \pm 0.59 & (D) \\
DESJ0540-2127 & 8 & RNA, ACT, RedM, SPTPol & 14.37 \pm 0.38 & SPT-CL J0540-2127 (m) \\
DESJ0545-2635 & 8 & RNA, BBNA & 13.37 \pm 0.29 & DESI-086.3066-26.5884 (z) \\
DESJ0548-3614 & 7 & RedM & 7.09 \pm 0.76 & DESI-087.1525-36.2427 (z) \\
DESJ0553-3342 & 6 & RNA, ACT, SPTPol & 11.41 \pm 2.17 & SPT-CL J0553-3342 (m), (r) \\
DESJ0557-4113 & 6 & ACT & 15.77 \pm 1.60 & (B) \\
DESJ0603-3558 & 9 & ACT & 16.35 \pm 0.95 & DESI-090.9854-35.9683 (z) \\
DESJ0604-3347 & 5 & RNA, RedM & 7.86 \pm 0.83 & \ldots \\
DESJ0610-5559A & 8 & RedM & 4.06 \pm 0.48 & (B) \\
DESJ0610-5559B & 8 & RedM & 9.22 \pm 0.37 & (B) \\
DESJ0611-5514A & 9 & RNA, RedM & 9.26 \pm 1.42 & (B) \\
DESJ0611-5514B & 9 & RNA, RedM & 8.06 \pm 1.38 & (B) \\
DESJ0612-3920 & 3 & RNA, ACT, RedM & 3.59 \pm 0.44 & (D) \\
DESJ2011-5228 & 10 & RNA, ACT & 13.04 \pm 4.23 & SPT-CL J2011-5228 (q), (B) \\
DESJ2011-5725 & 3 & ACT, BBNA & 13.62 \pm 1.39 & \ldots \\
DESJ2019-5642 & 6 & ACT & 17.39 \pm 1.76 & \ldots \\
DESJ2022-6032 & 4 & RNA, RedM missed & 7.66 \pm 0.32 & \ldots \\
DESJ2023-5535 & 3 & ACT, RedM missed & 7.44 \pm 0.79 & \ldots \\
DESJ2025-5117 & 4 & RNA, ACT & 7.29 \pm 0.77 & (B) \\
DESJ2028-4316 & 3 & RNA, RedM missed & 4.45 \pm 0.52 & \ldots \\
DESJ2031-4037 & 6 & ACT & 23.88 \pm 2.40 & SMACS J2031.8-4036 (k) \\
DESJ2046-6146 & 3 & RNA, RedM missed & 6.64 \pm 0.71 & \ldots \\
DESJ2106-4411 & 3 & RedM & 4.79 \pm 1.15 & (D) \\
DESJ2111-0114 & 9 & ACT & 10.73 \pm 0.60 & ClG J2111-0115 (c), (e), (B) \\
DESJ2112-4801 & 8 & RNA, RedM & 6.31 \pm 0.68 & \ldots \\
DESJ2122-0059 & 8 & RedM & 3.00 \pm 0.40 & (B) \\
DESJ2124-6125 & 9 & RNA, RedM & 3.70 \pm 0.45 & \ldots \\
DESJ2134-4238 & 3 & ACT, RedM missed & 31.98 \pm 3.21 & \ldots \\
DESJ2138-6007 & 10 & RedM & 22.45 \pm 7.61 & \ldots \\
DESJ2139-4251 & 6 & RNA, RedM & 2.83 \pm 0.39 & (D) \\
DESJ2144-4323 & 3 & RedM & 7.73 \pm 0.82 & \ldots \\
DESJ2145-5644 & 5 & ACT, RedM & 28.95 \pm 0.66 & \ldots \\
DESJ2152-5555 & 6 & BBNA & 3.21 \pm 0.42 & \ldots \\
DESJ2154-4604 & 6 & RedM & 3.91 \pm 0.47 & \ldots \\
DESJ2157-5700 & 4 & RedM & 4.61 \pm 0.53 & (B) \\
DESJ2159-6245 & 6 & RNA, RedM missed & 5.90 \pm 0.65 & \ldots \\
DESJ2200-4128 & 6 & ACT, RedM & 1.98 \pm 0.33 & (D) \\
DESJ2201-5956 & 3 & ACT, BBNA & 19.18 \pm 1.94 & \ldots \\
DESJ2208-0124 & 4 & BNA & 3.02 \pm 0.40 & (B) \\
DESJ2219-4348 & 7 & RedM & 3.15 \pm 0.36 & DES J221912-434835 (A), (D) \\
DESJ2223-6329 & 7 & RedM & 4.83 \pm 0.62 & \ldots \\
DESJ2232-5959 & 4 & RNA, ACT & 8.56 \pm 0.37 & (D) \\
DESJ2232-5807 & 7 & RedM & 8.11 \pm 0.85 & (B) \\
DESJ2233-0104 & 4 & ACT & 5.12 \pm 0.48 & HSCJ223316-010409 (w) \\
DESJ2240-5245 & 5 & ACT, RedM & 6.17 \pm 0.67 & (B) \\
DESJ2248-4431 & 7 & RNA, ACT & 32.51 \pm 0.40 & Abell S1063 (n), (B), (D) \\
DESJ2254-4620 & 5 & ACT, RedM missed & 33.45 \pm 3.26 & (B) \\
DESJ2300-5820 & 5 & RNA, RedM missed & 8.06 \pm 0.35 & (B) \\
DESJ2308-0211 & 7 & RNA, RedM & 35.98 \pm 3.61 & Abell 2537 (s), (z), (D) \\
DESJ2311-6307 & 5 & BNA & 2.86 \pm 0.39 & \ldots \\
DESJ2321-4630 & 9 & BNA & 3.13 \pm 0.32 & DES J232128-463049 (A), (B), (D) \\
DESJ2322-6409 & 9 & RedM & 9.19 \pm 0.58 & (D) \\
DESJ2324-4944 & 7 & RedM & 5.08 \pm 0.29 & (B) \\
DESJ2325-4111 & 7 & ACT, RedM & 42.37 \pm 0.33 & (B), (D) \\
DESJ2331-5051 & 7 & ACT & 16.84 \pm 0.30 & \ldots \\
DESJ2332-0152 & 5 & BNA & 3.40 \pm 0.90 & \ldots \\
DESJ2335-4209 & 5 & RedM & 4.25 \pm 0.50 & \ldots \\
DESJ2335-5152 & 7 & RNA, RedM & 3.64 \pm 0.50 & DES J233551-515217 (A), (B), (D) \\
DESJ2336-5352 & 7 & ACT, RedM & 6.33 \pm 2.03 & DES J2336-5352 (o), (B), (D) \\
DESJ2341-5716 & 7 & RedM & 8.55 \pm 0.29 & DESI-355.2727-57.2679 (z) \\
DESJ2342-4652 & 5 & RedM & 7.27 \pm 0.77 & \ldots \\
DESJ2343-6039 & 6 & RedM & 11.11 \pm 1.14 & \ldots \\
DESJ2347-6246 & 7 & RedM & 6.30 \pm 0.68 & DESI-356.7894-62.7765 (z) \\
DESJ2347-6245 & 3 & RedM & 10.34 \pm 1.07 & \ldots \\
DESJ2349-5113 & 9 & RNA, RedM & 4.51 \pm 0.70 & DES J234930-511339 (A), (B), (D) \\
DESJ2351-5452 & 10 & RNA, ACT, RedM & 7.43 \pm 1.20 & Elliot Arc (f), (B), (D) \\
DESJ2358-6125 & 8 & RedM & 4.82 \pm 0.55 & DESI-359.7003-61.4330 (z) \\
DESJ2359+0208 & 9 & RNA, ACT, RedM & 8.88 \pm 0.80 & DESI-359.8897+02.1399 (v), (w) \\

    [1ex]
    \enddata

    \tablecomments{Properties of all systems presented in this work. Names, algorithms that detected the system, the visual inspection rank, average radius, and references to detections in other papers. Systems labeled ``RedM missed" were not identified in the RedM search, but were later found to match redMaPPer galaxy clusters. The algorithms are as follows: BNA = Blue Near Anything, BBNA = secondary Blue Near Anything search, RNA = Red Near Anything, TI = Tile Inspection,ACT = Atacama Cosmology Telescope cluster sample, RedM = redMaPPer year 3 cluster sample, SPTPol = SPTPol Extended Cluster Survey clusters.
    \textbf{References:} Names are only given for the earliest discovery. Names from \citet{diehlsvy1,Jacobs.Collett.ea2019_alt,Jacobs.Collett.ea2019a} are identical to the present work.
    $^a$\cite{Cabanac.Alard.ea2007}
$^b$\cite{Estrada.Annis.ea2007}
$^c$\cite{hennawi2008}
$^d$\cite{Wuyts.Barrientos.ea2010}
$^e$\cite{Bayliss.Hennawi.ea2011}
$^f$\cite{elliotarc}
$^g$\cite{Bayliss2012}
$^h$\cite{furlanetto2013}
$^i$\cite{Stark2013}
$^j$\cite{Jauzac.Clement.ea2014}
$^k$\cite{Richard.Patricio.ea2015}
$^l$\cite{Zitrin.Fabris.ea2015}
$^m$\cite{SPTBleem}
$^n$\cite{Caminha2016}
$^o$\cite{Nord.Buckley-Geer.ea2016}
$^p$\cite{Bayliss.Ruel.ea2016}
$^q$\cite{collett2017}
$^r$\cite{Ebeling2017}
$^s$\cite{Cerny.Sharon.ea2018}
$^t$\cite{DiegoSchmidt2018}
$^u$\cite{Petrillo.Tortora.ea2019}
$^v$\cite{Huang.Storfer.ea2020}
$^w$\cite{Jaelani2020}
$^x$\cite{Li.Napolitano.ea2020}
$^y$\cite{Canameras.Schuldt.ea2020}
$^z$\cite{Huang.Storfer.ea2021}
$^A$\cite{Rojas2021}
$^B$\cite{diehlsvy1}
$^C$\cite{Jacobs.Collett.ea2019_alt}
$^D$\cite{Jacobs.Collett.ea2019a}

    }
\end{deluxetable*}


\color{black}{}
\startlongtable
\begin{deluxetable*}{lccCC}
    \tablecaption{Names, Positions, Photometry, and Photometric Redshifts of Objects for Each Candidate Lensing System
        \label{tab:all_search_objects}}
    \tabletypesize{\scriptsize}

    \tablehead{
        \colhead{System Name} & \colhead{RA (J2000)} & \colhead{Dec (J2000)}
        & \colhead{Magnitudes (g, r, i, z, y)} & \colhead{z$_{\rm{photo}}$}
        \vspace{-0.08in}
        \\ \colhead{(Object Label)} & & & &
    }

    \newcounter{asteriskfootnote}
    \setcounter{asteriskfootnote}{1}
    \startdata
        DESJ0002-3332 (A) & 0.527213 & -33.544062 & ($22.67 \pm 0.08$, $21.00 \pm 0.02$, $20.35 \pm 0.02$, $20.02 \pm 0.04$, $19.92 \pm 0.08$) & $0.50 \pm 0.03$ \\
DESJ0002-3332 (B) & 0.527307 & -33.542925 & ($22.45 \pm 0.05$, $21.12 \pm 0.02$, $20.70 \pm 0.02$, $20.45 \pm 0.04$, $20.44 \pm 0.11$) & $0.48 \pm 0.08$ \\
DESJ0002-3332 (1) & 0.525724 & -33.543500 & ($22.12 \pm 0.07$, $21.05 \pm 0.04$, $20.50 \pm 0.04$, $20.27 \pm 0.07$, $20.54 \pm 0.23$) & $0.68 \pm 0.05$ \\
DESJ0002-3332 (2) & 0.526278 & -33.542441 & ($22.57 \pm 0.09$, $21.64 \pm 0.05$, $21.15 \pm 0.05$, $21.12 \pm 0.13$, $21.42 \pm 0.40$) & $0.67 \pm 0.06$ \\
\hline
DESJ0007-4434 (A) & 1.872055 & -44.579515 & ($20.42 \pm 0.01$, $18.73 \pm 0.00$, $18.01 \pm 0.00$, $17.62 \pm 0.01$, $17.45 \pm 0.01$) & $0.51 \pm 0.01$ \\
DESJ0007-4434 (1) & 1.870997 & -44.578933 & ($21.74 \pm 0.01$, $21.01 \pm 0.01$, $20.42 \pm 0.01$, $20.08 \pm 0.01$, $19.95 \pm 0.03$) & $0.73 \pm 0.09$ \\
\hline
DESJ0011-4614 (A) & 2.971390 & -46.239450 & ($20.18 \pm 0.01$, $19.14 \pm 0.00$, $18.46 \pm 0.00$, $18.00 \pm 0.01$, $17.84 \pm 0.01$)\textsuperscript{\fnsymbol{asteriskfootnote}} & $0.57 \pm 0.02$\textsuperscript{\fnsymbol{asteriskfootnote}} \\
DESJ0011-4614 (1) & 2.970150 & -46.238660 & \ldots & \ldots \\
DESJ0011-4614 (2) & 2.973640 & -46.241360 & \ldots & \ldots \\
\hline
DESJ0014-0057 (A) & 3.725464 & -0.952317 & ($21.14 \pm 0.03$, $19.24 \pm 0.01$, $18.45 \pm 0.01$, $18.04 \pm 0.01$, $17.94 \pm 0.02$) & $0.53 \pm 0.00$ \\
DESJ0014-0057 (1) & 3.728786 & -0.951020 & ($21.25 \pm 0.06$, $20.28 \pm 0.04$, $19.64 \pm 0.03$, $19.41 \pm 0.05$, $19.35 \pm 0.16$)\textsuperscript{\fnsymbol{asteriskfootnote}} & $0.54 \pm 0.19$\textsuperscript{\fnsymbol{asteriskfootnote}} \\
\hline
DESJ0021-5028 (A) & 5.452849 & -50.476040 & ($20.37 \pm 0.01$, $18.80 \pm 0.00$, $18.26 \pm 0.00$, $17.94 \pm 0.01$, $17.81 \pm 0.02$) & $0.37 \pm 0.02$ \\
DESJ0021-5028 (1) & 5.454124 & -50.475333 & ($21.65 \pm 0.01$, $20.94 \pm 0.01$, $20.63 \pm 0.02$, $20.37 \pm 0.02$, $20.34 \pm 0.08$) & $0.20 \pm 0.06$ \\
\hline
DESJ0025-3139 (A) & 6.292237 & -31.657593 & ($21.38 \pm 0.05$, $19.88 \pm 0.02$, $19.15 \pm 0.02$, $18.81 \pm 0.02$, $18.57 \pm 0.04$)\textsuperscript{\fnsymbol{asteriskfootnote}} & $0.56 \pm 0.03$\textsuperscript{\fnsymbol{asteriskfootnote}} \\
DESJ0025-3139 (1) & 6.291383 & -31.657199 & ($22.09 \pm 0.03$, $21.46 \pm 0.03$, $21.04 \pm 0.03$, $20.78 \pm 0.04$, $20.70 \pm 0.10$) & $0.22 \pm 0.15$ \\
\hline
DESJ0025-4946 (A) & 6.294487 & -49.774042 & ($22.23 \pm 0.06$, $20.66 \pm 0.02$, $19.58 \pm 0.01$, $19.19 \pm 0.02$, $19.04 \pm 0.04$)\textsuperscript{\fnsymbol{asteriskfootnote}} & $0.69 \pm 0.04$\textsuperscript{\fnsymbol{asteriskfootnote}} \\
DESJ0025-4946 (1) & 6.295470 & -49.774668 & ($22.42 \pm 0.06$, $21.61 \pm 0.04$, $21.03 \pm 0.04$, $20.70 \pm 0.06$, $21.08 \pm 0.24$) & $0.75 \pm 0.06$ \\
\hline
DESJ0026-4825 (A) & 7.049739 & -48.435200 & ($19.37 \pm 0.01$, $17.60 \pm 0.00$, $17.03 \pm 0.00$, $16.71 \pm 0.00$, $16.57 \pm 0.01$)\textsuperscript{\fnsymbol{asteriskfootnote}} & $0.38 \pm 0.01$\textsuperscript{\fnsymbol{asteriskfootnote}} \\
DESJ0026-4825 (1) & 7.051111 & -48.432314 & ($21.62 \pm 0.02$, $20.57 \pm 0.01$, $20.16 \pm 0.01$, $19.82 \pm 0.02$, $19.67 \pm 0.04$) & $0.45 \pm 0.13$ \\
DESJ0026-4825 (2) & 7.049190 & -48.432200 & \ldots & \ldots \\
\hline
DESJ0030-4507 (A) & 7.663571 & -45.117814 & ($20.88 \pm 0.01$, $19.19 \pm 0.00$, $18.62 \pm 0.00$, $18.28 \pm 0.01$, $18.17 \pm 0.02$) & $0.40 \pm 0.04$ \\
DESJ0030-4507 (1) & 7.661060 & -45.115780 & \ldots & \ldots \\
\hline
DESJ0031-3717 (A) & 7.779623 & -37.292908 & ($22.28 \pm 0.06$, $20.66 \pm 0.02$, $19.56 \pm 0.01$, $19.05 \pm 0.01$, $18.89 \pm 0.03$) & $0.74 \pm 0.03$ \\
DESJ0031-3717 (1) & 7.781746 & -37.293924 & ($22.84 \pm 0.07$, $22.47 \pm 0.07$, $22.18 \pm 0.09$, $21.91 \pm 0.13$, $21.35 \pm 0.25$) & $0.90 \pm 0.35$ \\
DESJ0031-3717 (2) & 7.780348 & -37.292305 & ($23.27 \pm 0.05$, $22.28 \pm 0.03$, $21.67 \pm 0.03$, $21.37 \pm 0.03$, $21.28 \pm 0.10$) & $0.62 \pm 0.11$ \\
DESJ0031-3717 (3) & 7.779335 & -37.291970 & ($22.72 \pm 0.06$, $21.92 \pm 0.04$, $21.51 \pm 0.05$, $21.12 \pm 0.06$, $20.86 \pm 0.16$) & $0.42 \pm 0.12$ \\
\hline
DESJ0032+0100 (A) & 8.073145 & 1.010437 & ($20.08 \pm 0.02$, $18.28 \pm 0.00$, $17.67 \pm 0.00$, $17.33 \pm 0.00$, $17.18 \pm 0.01$)\textsuperscript{\textdagger} & $0.39 \pm 0.00$\textsuperscript{\textdagger} \\
DESJ0032+0100 (1) & 8.072042 & 1.009085 & ($22.30 \pm 0.06$, $21.16 \pm 0.03$, $20.62 \pm 0.03$, $20.41 \pm 0.04$, $20.33 \pm 0.14$)\textsuperscript{\textdagger} & $0.23 \pm 0.06$\textsuperscript{\textdagger} \\
DESJ0032+0100 (2) & 8.071510 & 1.011570 & \ldots & \ldots \\
\hline
DESJ0034+0225 (A) & 8.617365 & 2.422945 & ($20.16 \pm 0.01$, $18.46 \pm 0.00$, $17.88 \pm 0.00$, $17.57 \pm 0.00$, $17.50 \pm 0.01$)\textsuperscript{\fnsymbol{asteriskfootnote}} & $0.38 \pm 0.03$\textsuperscript{\fnsymbol{asteriskfootnote}} \\
DESJ0034+0225 (1) & 8.615521 & 2.426298 & ($23.09 \pm 0.32$, $20.59 \pm 0.04$, $20.27 \pm 0.06$, $20.65 \pm 0.17$, $99.00 \pm 99.00$) & $0.66 \pm 0.71$ \\
DESJ0034+0225 (2) & 8.614283 & 2.422975 & ($21.93 \pm 0.06$, $20.79 \pm 0.03$, $20.45 \pm 0.04$, $20.41 \pm 0.08$, $20.48 \pm 0.24$)\textsuperscript{\fnsymbol{asteriskfootnote}} & $0.21 \pm 0.18$\textsuperscript{\fnsymbol{asteriskfootnote}} \\
DESJ0034+0225 (3) & 8.613881 & 2.420758 & ($22.11 \pm 0.05$, $21.66 \pm 0.05$, $21.59 \pm 0.08$, $21.57 \pm 0.17$, $22.62 \pm 1.32$) & $0.72 \pm 0.67$ \\
\hline
DESJ0035-2526 (A) & 8.779998 & -25.449930 & ($22.47 \pm 0.03$, $21.41 \pm 0.01$, $20.49 \pm 0.01$, $20.00 \pm 0.01$, $19.87 \pm 0.04$)\textsuperscript{\fnsymbol{asteriskfootnote}} & $0.64 \pm 0.51$\textsuperscript{\fnsymbol{asteriskfootnote}} \\
DESJ0035-2526 (1) & 8.780002 & -25.450581 & ($20.54 \pm 0.01$, $20.03 \pm 0.01$, $19.85 \pm 0.01$, $19.68 \pm 0.02$, $19.47 \pm 0.05$) & $0.05 \pm 0.05$ \\
\hline
DESJ0038-2155 (A) & 9.716928 & -21.924359 & ($21.15 \pm 0.02$, $19.84 \pm 0.01$, $19.42 \pm 0.01$, $19.15 \pm 0.01$, $18.99 \pm 0.03$) & $0.39 \pm 0.05$ \\
DESJ0038-2155 (1) & 9.717215 & -21.922871 & ($22.42 \pm 0.07$, $21.47 \pm 0.04$, $21.35 \pm 0.06$, $21.13 \pm 0.09$, $21.07 \pm 0.27$)\textsuperscript{\fnsymbol{asteriskfootnote}} & $0.11 \pm 0.00$\textsuperscript{\fnsymbol{asteriskfootnote}} \\
DESJ0038-2155 (2) & 9.715270 & -21.924870 & \ldots & \ldots \\
\hline
    \enddata

    \tablecomments{Only the first 13 systems are shown, data for all systems is available in machine readable format. Objects which did not have a corresponding detection in DES Y3 or DES Y6 data have no photometry or photometric redshift, represented as ellipses (\ldots). Photometry and photometric redshifts with (without) a dagger (\textdagger) are derived from DES Y3 (DES Y6) data. Photometry and photometric redshifts marked with (\fnsymbol{asteriskfootnote})
    have \texttt{FLAGS\_GOLD} $!= 0$, indicating a likely problem with the object's identification or photometry.
    }
\end{deluxetable*}

\begin{figure}[ht]
\begin{center}
\includegraphics[scale=0.26]{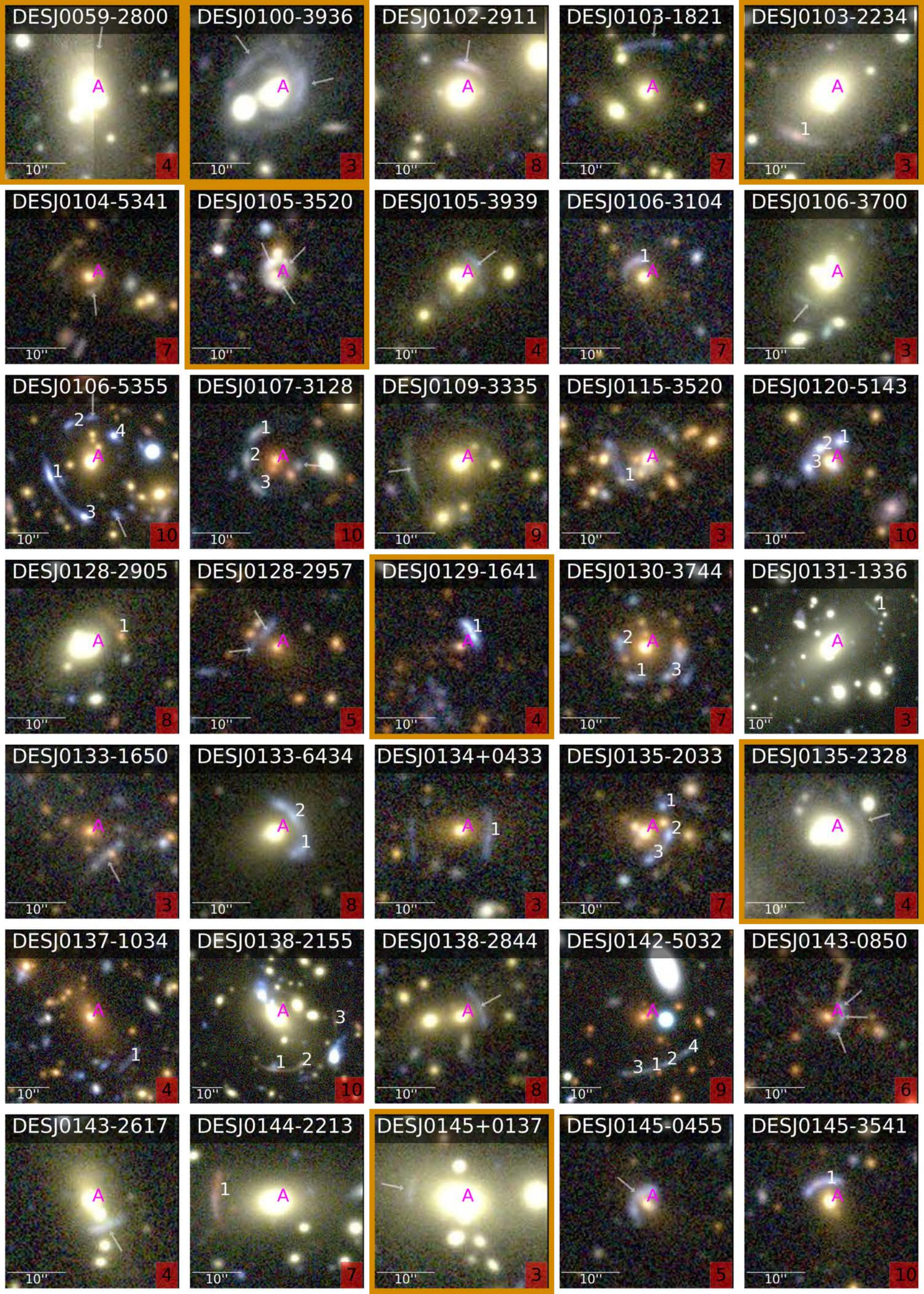}
\caption{Second page of SL systems with rank 3 or more.   Each cutout image has the visual inspection ranking displayed in a red box in the lower right hand corner. All images are oriented north up, east left.  Each cutout is dynamically sized to fit the SL system well within the cutout, with a minimum size of 30\arcsec \ $\times$ 30\arcsec.  A scale bar 10\arcsec \ long is displayed in the lower left hand corner.  New SL systems are outlined with a gold border, previously known systems with white. }
\end{center}
\end{figure}

\begin{figure}[ht]
\begin{center}
\includegraphics[scale=0.26]{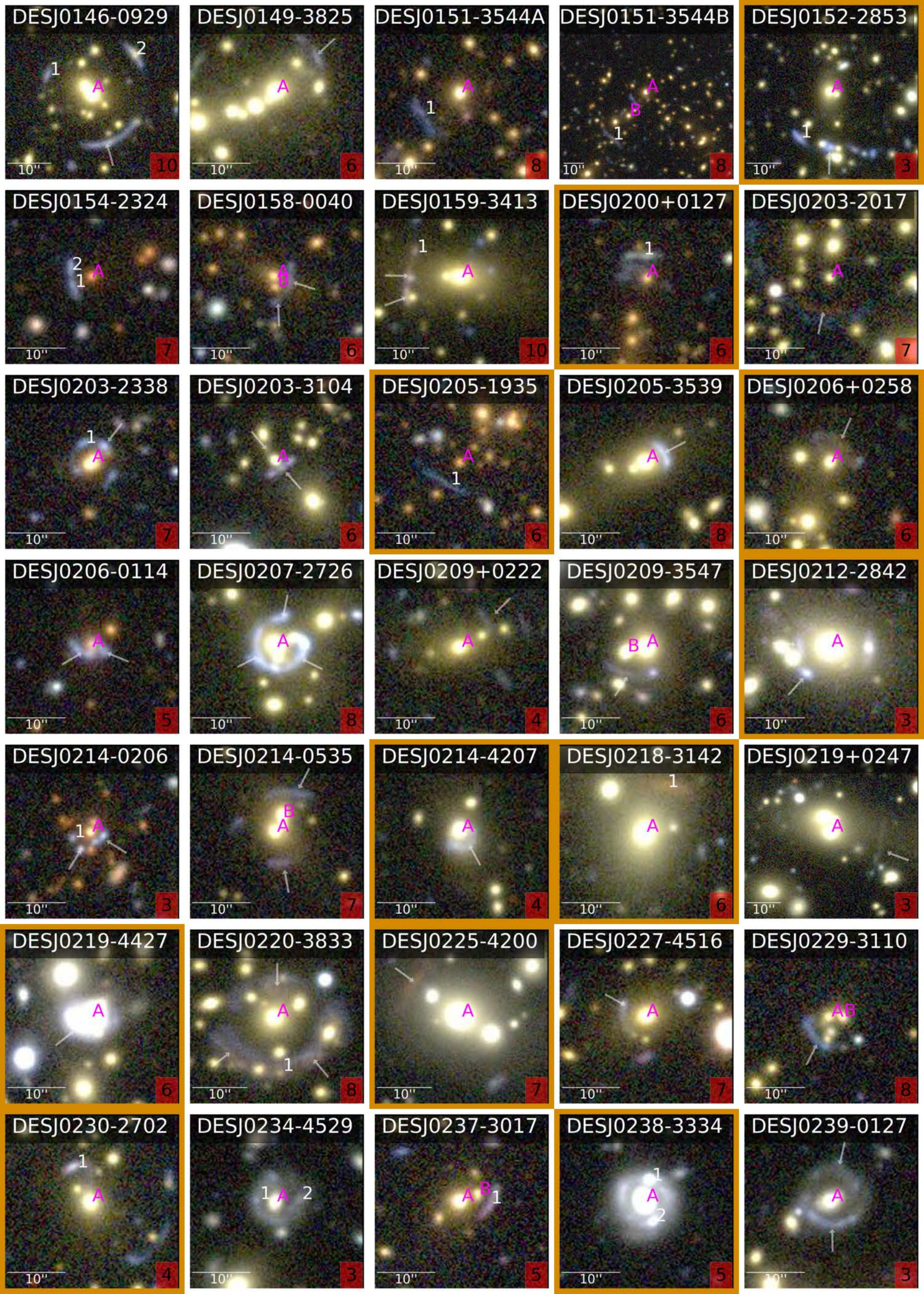}
\caption{Third page of SL systems with rank 3 or more.   Each cutout image has the visual inspection ranking displayed in a red box in the lower right hand corner. All images are oriented north up, east left.  Each cutout is dynamically sized to fit the SL system well within the cutout, with a minimum size of 30\arcsec \ $\times$ 30\arcsec.  A scale bar 10\arcsec \ long is displayed in the lower left hand corner.  New SL systems are outlined with a gold border, previously known systems with white. }
\end{center}
\end{figure}

\begin{figure}[ht]
\begin{center}
\includegraphics[scale=0.26]{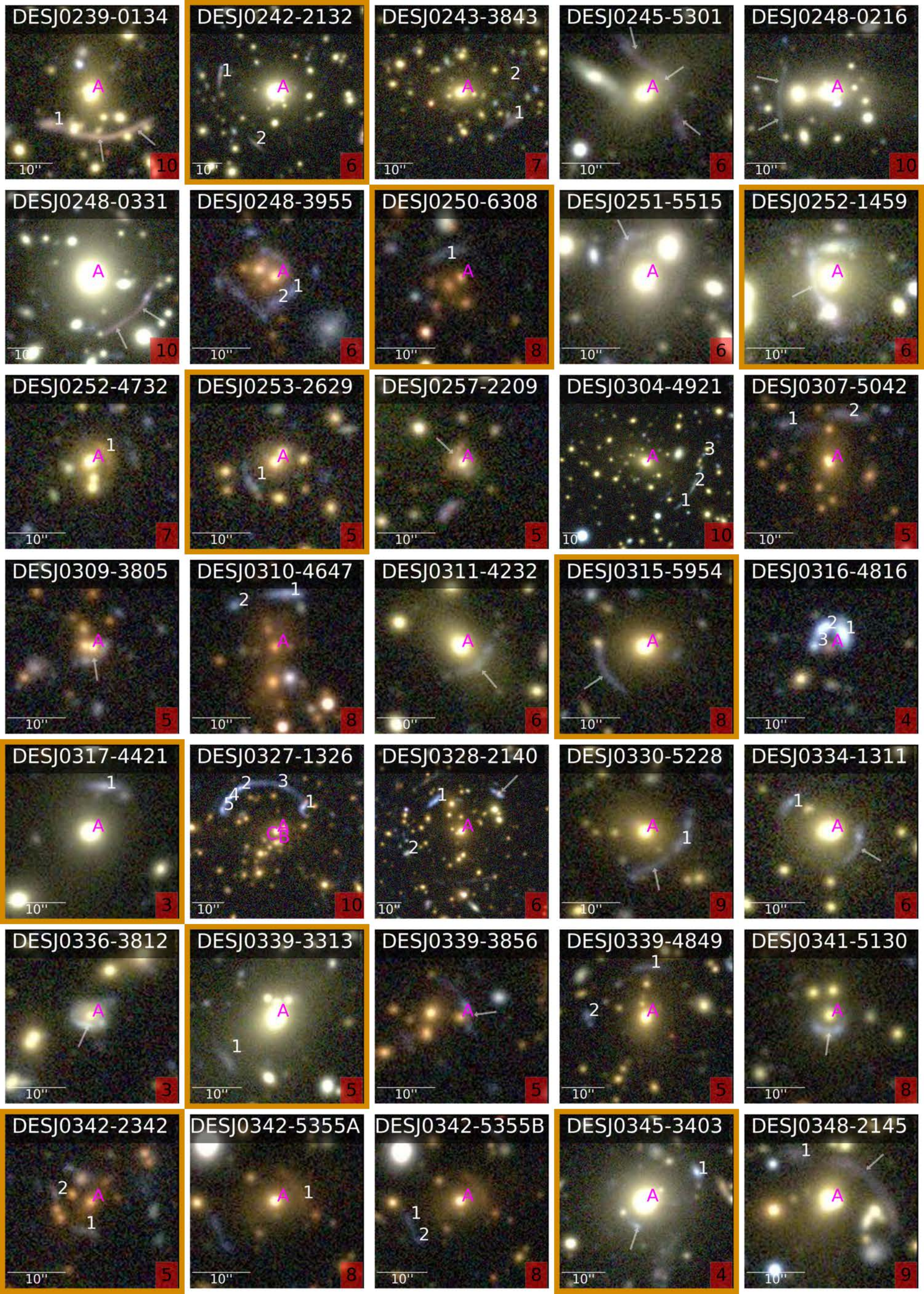}
\caption{Fourth page of SL systems with rank 3 or more.   Each cutout image has the visual inspection ranking displayed in a red box in the lower right hand corner. All images are oriented north up, east left.  Each cutout is dynamically sized to fit the SL system well within the cutout, with a minimum size of 30\arcsec \ $\times$ 30\arcsec.  A scale bar 10\arcsec \ long is displayed in the lower left hand corner.  New SL systems are outlined with a gold border, previously known systems with white.}
\end{center}
\end{figure}

\begin{figure}[ht]
\begin{center}
\includegraphics[scale=0.26]{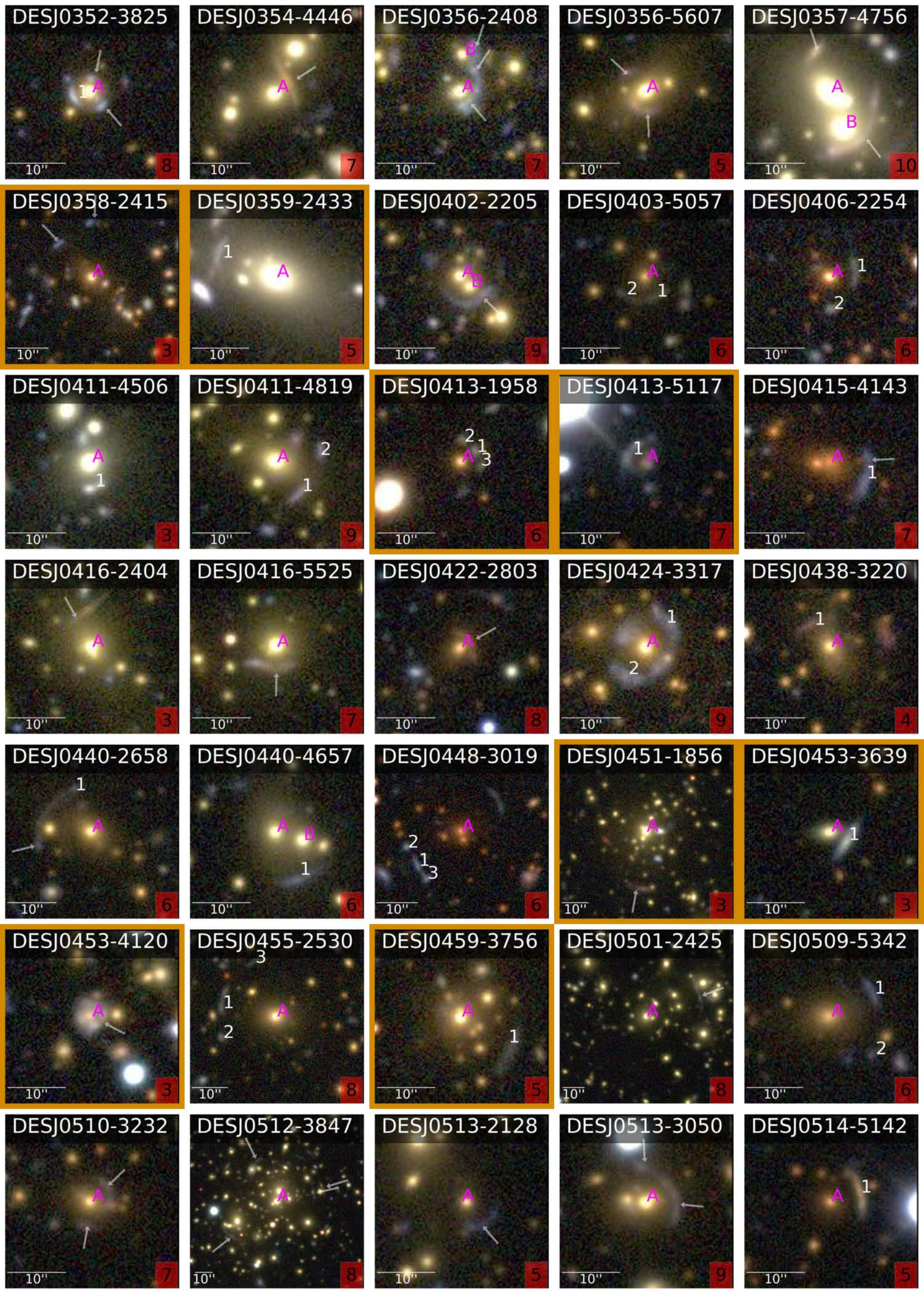}
\caption{Fifth page of SL systems with rank 3 or more.   Each cutout image has the visual inspection ranking displayed in a red box in the lower right hand corner. All images are oriented north up, east left.  Each cutout is dynamically sized to fit the SL system well within the cutout, with a minimum size of 30\arcsec \ $\times$ 30\arcsec.  A scale bar 10\arcsec \ long is displayed in the lower left hand corner.  New SL systems are outlined with a gold border, previously known systems with white.}
\end{center}
\end{figure}

\begin{figure}[ht]
\begin{center}
\includegraphics[scale=0.26]{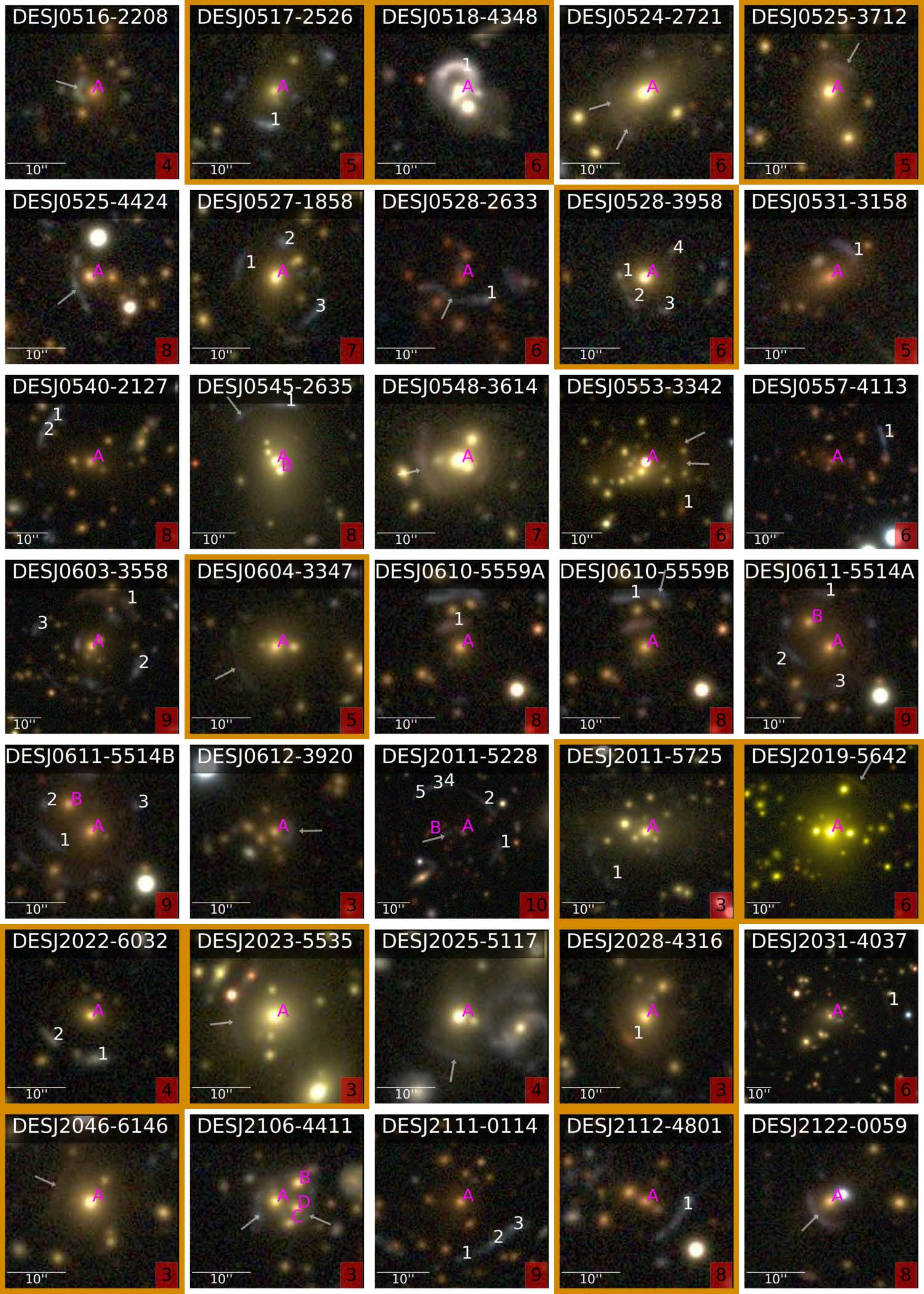}
\caption{Sixth page of SL systems with rank 3 or more.   Each cutout image has the visual inspection ranking displayed in a red box in the lower right hand corner. All images are oriented north up, east left.  Each cutout is dynamically sized to fit the SL system well within the cutout, with a minimum size of 30\arcsec \ $\times$ 30\arcsec.  A scale bar 10\arcsec \ long is displayed in the lower left hand corner.  New SL systems are outlined with a gold border, previously known systems with white. }
\end{center}
\end{figure}

\begin{figure}[ht]
\begin{center}
\includegraphics[scale=0.26]{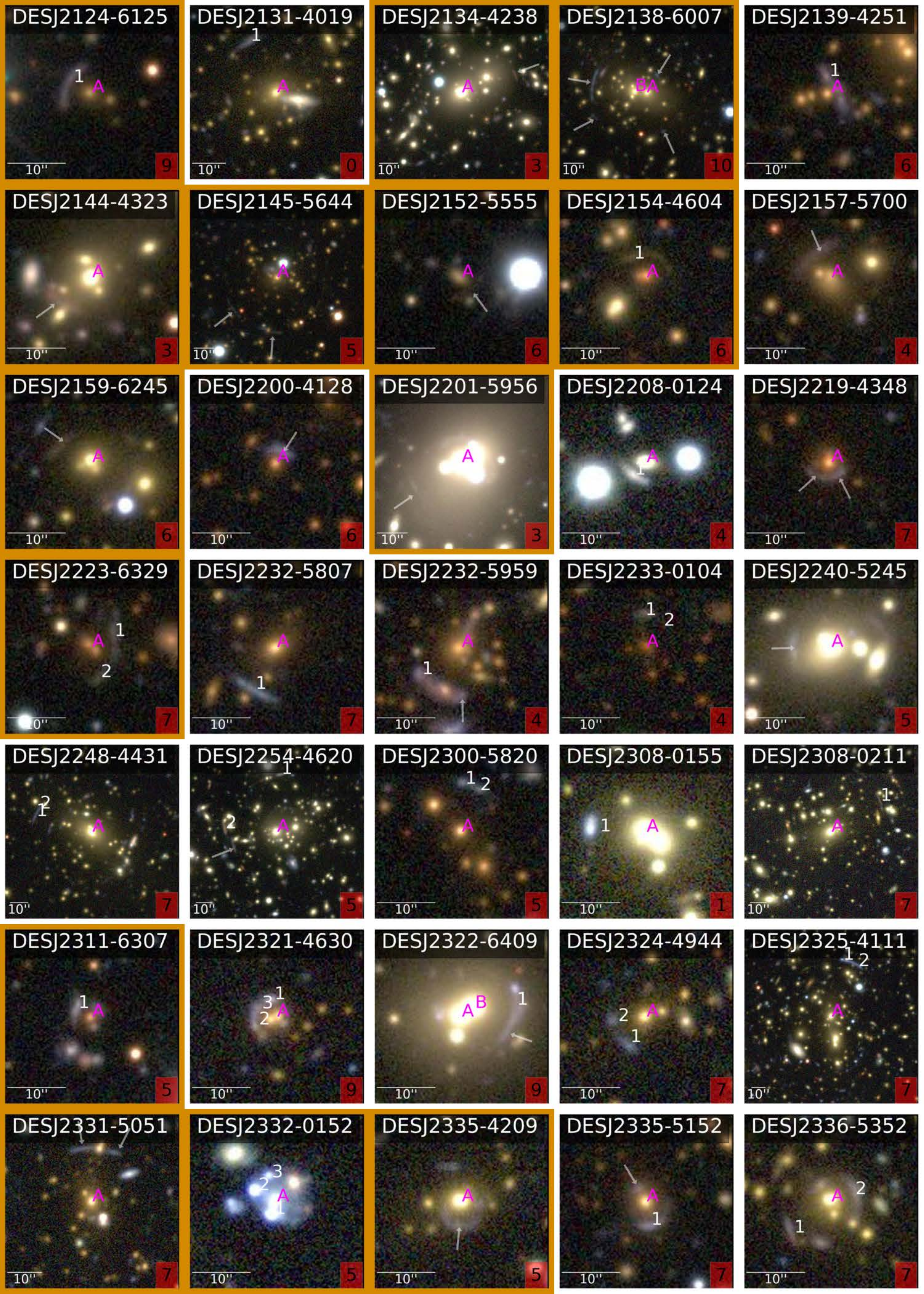}
\caption{Seventh page of SL systems with rank 3 or more.   Each cutout image has the visual inspection ranking displayed in a red box in the lower right hand corner. All images are oriented north up, east left.  Each cutout is dynamically sized to fit the SL system well within the cutout, with a minimum size of 30\arcsec \ $\times$ 30\arcsec.  A scale bar 10\arcsec \ long is displayed in the lower left hand corner.  New SL systems are outlined with a gold border, previously known systems with white. }
\end{center}
\end{figure}

\subsection{Notable Systems}

Here we highlight some notable systems.  The newly identified system DESJ2138-6007 
is particularly spectacular. It contains a long blue arc with radius nearly 30\arcsec\,and several possible counterimages. The system is centered on a redMaPPer Y3 galaxy cluster of richness $152 \pm 5$.
Four SL systems appear to contain two sources, one red and one blue. In each case the system is listed twice in Tables~\ref{tab:all_search_systems} and~\ref{tab:all_search_objects}, with suffixes A and B for the two distinct sources. These systems are DESJ0151-3544, DESJ0342-5355, DESJ0610-5559, and DESJ0611-5514, which have Y3
redMaPPer richnesses of $73 \pm 3$, $109 \pm 4$, $48 \pm 4$, and $44 \pm 4$, respectively. System DESJ0151-3544 was previously discovered in SPT follow up \citep{SPTBleem}, and systems DESJ0342-5355 and DESJ0610-5559 were identified and noted as double source SL systems in \cite{diehlsvy1}. The final one, DESJ0611-5514, is new.

System DESJ0043-2037 is a large, unrelaxed, potential merging galaxy cluster which appears to have two red arcs near its center. In this work, it is presented as two systems with suffixes A and B, centered on two separate luminous galaxies. It is a complex system which merits further study.

 Higher resolution imaging of other systems presented here might expose  additional remarkable features, such as central images, radial arcs, or substructure lensing in some of these systems. These features would provide significant information about the matter distribution in the lens systems.

 
\section{Summary and Conclusion}


We report the results of eight searches for strong gravitational lens candidates in the full 5,000 sq. deg. footprint of the DES imaging data. We searched the DES catalogs for spatial matches of potential lens and source candidates using the ``Blue Near Anything" and ``Red Near Anything" algorithms. We searched the positions of known ACT, SPTPol, and DES redMaPPer galaxy clusters. The scanning of wide-area tiles for the DES data quality effort led to serendipitous discovery of several SL systems.  For all of the searches we produced a short list of candidates and then evaluated cutouts to identify the most promising systems based on color and morphology.
We then assigned those a rank that quantifies our confidence that the system is a potential strong gravitational lens. Of the 247 systems that we found, 81 are presented for the first time. Of the newly discovered systems, at least one was uniquely found by each of our searches. For each system we provide the position, the magnitudes, and photometric properties of the lens and source objects, and the  distance (radius) of the source(s) from the lens center.
Some of these are striking systems with giant arcs.  Some have red-colored sources. Four have both blue and red candidate sources at differing distances from the candidate lens.

Each category of search (RNA, BNA, galaxy cluster, and tile inspection) provided unique newly-discovered systems, as shown in Figure~\ref{fig:systems-venn-diagram}. These varied search strategies serve as effective complements to each other, and each contributed new systems to the final catalog.







These SL systems tend to be galaxy group and cluster scale candidates, particularly those from the searches of galaxy clusters described in Section~\ref{sec3.4}. Therefore, we expect it will be a useful and valuable training and validation set for future automated searches of galaxy clusters in new fields and at lower richness range.  At the same time, this catalog also highlights the need for, and importance of, crowdsourced or automated lens modeling techniques~\citep{birrer2015,kung} being developed.\footnote{https://github.com/DES-SL/EasyLens and http://linan7788626.github.io/pages/Hoopla/index.html}

While a majority of the systems discovered in these searches were previously identified (166 of 247), it is clear that many undiscovered systems exist in current data sets. Predictions for the number of lenses in a survey depend on depth and area, but estimates for the full Dark Energy Survey range around a few thousand \citep{om2010}. In the near future, they are expected to expand to more than a hundred thousand in LSST data \citep{collett2015}. Fortunately, there will be no shortage of high quality strong lens systems for follow up and detailed study.

\color{black}
\section{Acknowledgements}

{\color{black}  This research has made use of the NASA/IPAC Extragalactic Database (NED) which is operated by the California Institute of Technology, under contract with the National Aeronautics and Space Administration.

Some figures in this work were generated with matplotlib \citep{matplotlib}.}

Funding for the DES Projects has been provided by the U.S. Department of Energy, the U.S. National Science Foundation, the Ministry of Science and Education of Spain, the Science and Technology Facilities Council of the United Kingdom, the Higher Education Funding Council for England, the National Center for Supercomputing  Applications at the University of Illinois at Urbana-Champaign, the Kavli Institute of Cosmological Physics at the University of Chicago,  the Center for Cosmology and Astro-Particle Physics at The Ohio State University, the Mitchell Institute for Fundamental Physics and Astronomy at Texas A\&M University, Financiadora de Estudos e Projetos, Funda{\c c}{\~a}o Carlos Chagas Filho de Amparo {\`a} Pesquisa do Estado do Rio de Janeiro, Conselho Nacional de Desenvolvimento Cient{\'i}fico e Tecnol{\'o}gico and the Minist{\'e}rio da Ci{\^e}ncia, Tecnologia e Inova{\c c}{\~a}o, the Deutsche Forschungsgemeinschaft and the Collaborating Institutions in the Dark Energy Survey.

The Collaborating Institutions are Argonne National Laboratory, the University of California at Santa Cruz, the University of Cambridge, Centro de Investigaciones Energ{\'e}ticas, Medioambientales y Tecnol{\'o}gicas-Madrid, the University of Chicago, University College London, the DES-Brazil Consortium, the University of Edinburgh, the Eidgen{\"o}ssische Technische Hochschule (ETH) Z{\"u}rich, Fermi National Accelerator Laboratory, the University of Illinois at Urbana-Champaign, the Institut de Ci{\`e}ncies de l'Espai (IEEC/CSIC), the Institut de F{\'i}sica d'Altes Energies, Lawrence Berkeley National Laboratory, the Ludwig-Maximilians Universit{\"a}t M{\"u}nchen and the associated Excellence Cluster Universe,  the University of Michigan, NSF's NOIRLab, the University of Nottingham, The Ohio State University, the University of Pennsylvania, the University of Portsmouth, SLAC National Accelerator Laboratory, Stanford University, the University of Sussex, Texas A\&M University, and the OzDES Membership Consortium.

Based in part on observations at Cerro Tololo Inter-American Observatory at NSF's NOIRLab (NOIRLab Prop. ID 2012B-0001; PI: J. Frieman), which is managed by the Association of Universities for Research in Astronomy (AURA) under a cooperative agreement with the National Science Foundation.

The DES data management system is supported by the National Science Foundation under Grant Numbers AST-1138766 and AST-1536171. The DES participants from Spanish institutions are partially supported by MICINN under grants ESP2017-89838, PGC2018-094773, PGC2018-102021, SEV-2016-0588, SEV-2016-0597, and MDM-2015-0509, some of which include ERDF funds from the European Union. IFAE is partially funded by the CERCA program of the Generalitat de Catalunya. Research leading to these results has received funding from the European Research Council under the European Union's Seventh Framework Program (FP7/2007-2013) including ERC grant agreements 240672, 291329, and 306478. We  acknowledge support from the Brazilian Instituto Nacional de Ci\^encia e Tecnologia (INCT) do e-Universo (CNPq grant 465376/2014-2).

This manuscript has been authored by Fermi Research Alliance, LLC under Contract No. DE-AC02-07CH11359 with the U.S. Department of Energy, Office of Science, Office of High Energy Physics.






\nocite{*}
\bibliography{main}{} 
\bibliographystyle{aasjournal_new}   




\end{document}